\documentclass{jltp}
\usepackage{amsmath,amssymb}
\usepackage{latexsym}

\newcommand{\comm}[2]{\left[#1,#2\right]}

\newcommand{\half}{$\frac{1}{2}$ }
\newcommand{\vac}{\left|\,0\,\right\rangle}
\newcommand{\ket}[1]{\left|#1\right\rangle}
\newcommand{\bra}[1]{\left\langle#1\right|}
\newcommand{\braket}[1]{\left\langle#1\right\rangle}
\newcommand{\up}{\uparrow}
\newcommand{\dw}{\downarrow}
\def\bd{\begin{displaymath}}
\def\ed{\end{displaymath}}
\def\beq{\begin{equation}}
\def\eeq{\end{equation}}
\def\bea{\begin{eqnarray}}
\def\eea{\end{eqnarray}}
\def\bar{\begin{array}{ccc}}
\def\ear{\end{array}}
\def\bi{\begin{itemize}}
\def\ei{\end{itemize}}
\def\bn{\begin{enumerate}}
\def\en{\end{enumerate}}

\def\ie{i.e.,\ }

\title{{\boldmath $S=1$} Spin Liquids: Broken Discrete Symmetries
  Restored\thanks{\rm Dedicated to Peter W\"olfle on the occasion of
    his 60$^{\rm th}$ birthday.}}

\author{Martin Greiter}

\address{Institut f\"ur Theorie der Kondensierten Materie,\\
  Universit\"at Karlsruhe, Postfach 6980, D-76128 Karlsruhe}

\runninghead{M.~Greiter}{{\boldmath $S=1$} spin liquids: broken
    discrete symmetries restored}

\begin{document}

\maketitle

\begin{abstract}
  We introduce a novel kind of spin liquid, the spin 1 chirality
  liquid, which provides a generic paradigm for a disordered spin 1
  antiferromagnet in two dimensions.  It supports spinon and holon
  excitations, which carry a chirality quantum number.  These
  excitations obey half-fermi statistics.  The sign of the statistical
  parameter is determined by the chirality.  The spin 1 chirality
  liquid is the first example of a two dimensional quantum state which
  supports excitations with fractional statistics but does not violate
  P or T.\\[-3pt]

PACS numbers: 05.30.Pr, 71.10.-w, 75.10.-b, 75.90.+w.
\end{abstract}

\section{INTRODUCTION}
The key thought of this article is easily expressed.  Whenever we have 
a spin liquid (or valence bond solid) state with spin-$s$ on a lattice,
we can obtain another spin liquid with spin-$ms$, where $m$ is an integer,
by combining $m$ copies of such a state and projecting out the spin-$ms$
representation contained in
\begin{equation}
\underbrace{{\bf s}\otimes{\bf s}\otimes\ldots \otimes {\bf s}}_{m}
={\bf ms}\oplus m\,{\bf (m-1)s}\oplus\hbox{\ldots}
\label{eq:rep}
\end{equation}
at each site.  This is particularly useful if the spin-$s$ state
(spontaneously) violates one or many discrete symmetries, which we 
desire not to be violated for the spin-$ms$ state we construct.  We then only have to 
combine one copy of each of the ``degenerate'' spin-$s$ states.  If there 
are now $m$ of them, we will obtain a state with spin-$ms$ which does not 
violate any of these symmetries.  As we shall see below, the projection
can be formulated very elegantly using Schwinger bosons\cite{schw}.

Affleck, Kennedy, Lieb and Tasaki (AKLT)\cite{aklt} have constructed a
series of states by availing themselves of this principle for valence bond
solids, which discretely violate various lattice symmetries depending
on lattice type and dimension.  In this article, I advocate that the
principle may have other interesting applications, and introduce a
particular and generically new example of a two dimensional spin 1
liquid, which I call the {\it spin 1 chirality liquid}.  This liquid
supports charge neutral spin-$\frac{1}{2}$ spinon (and spinless
charge $1$ holon) excitations, which carry a chirality quantum
number.  This number can be $\chi=+1$ or $-1$.  The excitations obey
half-fermi statistics, both in the sense of Haldane's generalized
exclusion principle\cite{excl} and in the sense of the
Berry's phases encountered by the state as particles
adiabatically encircle each other\cite{wilc}.  The chirality quantum
number determines the sign of this phase.  This liquid is the
first example of a state with fractional statistics excitations 
in more than one space dimension which does not
violate the discrete symmetries parity (P) and time reversal (T).

The spin 1 chirality liquid is, at least at the time of Peter
W\"olfle's 60$^{\rm th}$ 
birthday, mainly of conceptual importance.  As alluded
to above, it is rather straightforward to formulate it.  What is not so
easily explained is the conceptual importance, the motivation.  Why
are we interested in spin liquids in dimensions larger than one?  
Why should we believe
that they generically support spinon excitations, which carry spin
\half and no charge?  Why do I put forward the hypothesis that spinons
in any dimension must obey half-fermi statistics?$\,$\cite{sutwo}

If the reader agrees with my suggested answers to these questions, he
or she will be quick to realize that spin liquids in two space
dimensions must either violate P and T (as it has been advocated by
R.B.\ Laughlin\cite{chiral}) or support spinon (and holon) 
excitations,
which carry a chirality quantum number.  (The individual excitations
then still violate P and T, but this is not any more significant then
the P and T violation of right and left moving fermions in one
dimension: one can always use the complex eigenfunctions of any
supposed real Hamiltonian $H$ to construct real ones, and hence
combine them into simultaneous eigenfunctions of $H$, $P$, and $T$.)

In this article, I present the first spin liquid of this second
kind.  It is a spin liquid with spin $s=1$.  By now, I have also
succeeded in constructing a P and T invariant spin liquid for
$s=\frac{1}{2}$, which I believe to be relevant to the problem of
CuO-superconductivity\cite{else}.  The construction of a generically
invariant spin liquid is, however, only simple in the $s=1$ case
discussed here.  It is one of the more inspired applications of the
simple projection method outlined above, and at the same time the
simplest paradigm for a sophisticated principle---the principle of
fractional quantization and emerging particles without P or T violation 
in two-dimensional antiferromagnets---known today.

In the following three sections, I will illustrate the general
projection principle with the simplest example, the spin 1 AKLT chain,
which is obtained by combining two dimer or Majumdar-Gosh chains\cite{mg}.
In sections \ref{s:hs} and \ref{s:frac}, I will introduce the concepts
of Jastrow-type wave functions and spinon excitations in
antiferromagnetically correlated spin liquids in a brief review of the
Haldane-Shastry model\cite{hs}, and illustrate why I believe that
spinons must obey a fractional exclusion principle or
fractional statistics, regardless of the model specifics including
the dimension.  In section \ref{s:chiral}, I will review the chiral
spin liquid state\cite{chiral,wwz,zou}, which supports spinon
excitations with half-fermi statistics, but also violates P and T.  I
will introduce the spin 1 chirality liquid, which does not violate any
symmetry, in section \ref{s:cl}\ \  In section \ref{s:num},
I will present a modest amount of numerical work on spin 1 systems.
Finally, I will address the question of spinon confinement in the spin
1 chirality liquid in section \ref{s:con}

\section{THE MAJUMDAR-GOSH MODEL}
\label{s:mg}

Majumdar and Gosh\cite{mg} noticed in 1967 that on a linear spin 
$S=\frac{1}{2}$ chain with an even number of sites, the two 
valence bond solid or dimer states 
\begin{eqnarray}
\big|\psi^{\scriptscriptstyle\textrm{MG}}_{\textrm{even}\rule{0pt}{5pt}\atop 
\textrm{(odd)}}\big\rangle &=&
\prod_{i\ \textrm{even}\atop (i\ \textrm{odd})}  
(c^\dagger_{i\up} c^\dagger_{i+1\dw} - c^\dagger_{i\dw} c^\dagger_{i+1\up}) 
\vac =\nonumber \\
&=&\left\{ \begin{array}{lc}
\ket{\hbox{\begin{picture}(132,8)(-4,-3)
\put(13,0){\makebox(0,0){\rule{10.pt}{ 0.8pt}}}
\put(41,0){\makebox(0,0){\rule{10.pt}{ 0.8pt}}}
\put(69,0){\makebox(0,0){\rule{10.pt}{ 0.8pt}}}
\put(97,0){\makebox(0,0){\rule{10.pt}{ 0.8pt}}}
\put(120,0){\makebox(0,0)[l]{\rule{4.pt}{ 0.8pt}}}
\put(6,0){\circle{4}}
\put(20,0){\circle{4}}
\put(34,0){\circle{4}}
\put(48,0){\circle{4}}
\put(62,0){\circle{4}}
\put(76,0){\circle{4}}
\put(90,0){\circle{4}}
\put(104,0){\circle{4}}
\put(118,0){\circle{4}}
\end{picture}}} 
&\quad\textrm{``even''}\\
\ket{\hbox{\begin{picture}(132,8)(-4,-3)
\put(4,0){\makebox(0,0)[r]{\rule{4.pt}{ 0.8pt}}}
\put(27,0){\makebox(0,0){\rule{10.pt}{ 0.8pt}}}
\put(55,0){\makebox(0,0){\rule{10.pt}{ 0.8pt}}}
\put(83,0){\makebox(0,0){\rule{10.pt}{ 0.8pt}}}
\put(111,0){\makebox(0,0){\rule{10.pt}{ 0.8pt}}}
\put(6,0){\circle{4}}
\put(20,0){\circle{4}}
\put(34,0){\circle{4}}
\put(48,0){\circle{4}}
\put(62,0){\circle{4}}
\put(76,0){\circle{4}}
\put(90,0){\circle{4}}
\put(104,0){\circle{4}}
\put(118,0){\circle{4}}
\end{picture}}} 
&\quad\textrm{``odd''}\rule{0pt}{15pt}
\end{array}\right.\rule{0pt}{25pt}
\label{e:psimg}
\end{eqnarray}
where the product runs over all even sites $i$ for one state and over all odd
sites for the other, 
are exact zero energy ground states of the parent Hamiltonian
\begin{equation}
H^{\scriptscriptstyle\textrm{MG}} = 
\sum_i \left(\boldsymbol{S}_i \boldsymbol{S}_{i+1} +
\frac{1}{2}\boldsymbol{S}_i \boldsymbol{S}_{i+2} +\frac{3}{8}\right). 
\label{e:hmg}
\end{equation}
The proof is exceedingly simple.  We rewrite
\begin{equation} 
H^{\scriptscriptstyle\textrm{MG}}=
\frac{1}{4}\sum_i H_i \quad\hbox{with} \quad
H_i =\left(\boldsymbol{S}_i +\boldsymbol{S}_{i+1} 
+\boldsymbol{S}_{i+2}\right)^2 -\frac{3}{4}.
\end{equation}
Clearly, any state in which the total spin of three neighboring spins 
is $\frac{1}{2}$ will be annihilated by $H_i$.  (The total spin can only be 
$\frac{3}{2}$ or $\frac{1}{2}$, as 
$\bf\frac{1}{2}\otimes\frac{1}{2}\otimes\frac{1}{2}=
\frac{3}{2}\oplus\frac{1}{2}\oplus\frac{1}{2}$.)
In the dimer states
above, this is always the case as two of the three neighboring spins are in 
a singlet configuration, and 
$\bf 0\otimes\frac{1}{2}=\frac{1}{2}$.  
Graphically, we may express this as 
\begin{equation}
H_i \ket{\hbox{\begin{picture}(40,8)(0,-3)
\put(13,0){\makebox(0,0){\rule{10.pt}{ 0.8pt}}}
\put(6,0){\circle{4}}
\put(20,0){\circle{4}}
\put(34,0){\circle{4}}
\end{picture}}}
= H_i \ket{\hbox{\begin{picture}(40,8)(0,-3)
\put(27,0){\makebox(0,0){\rule{10.pt}{ 0.8pt}}}
\put(6,0){\circle{4}}
\put(20,0){\circle{4}}
\put(34,0){\circle{4}}
\end{picture}}}=0.
\end{equation}
 As $H_i$ is positive definite, the two zero energy eigenstates of
$H_{\textrm{MG}}$ are also ground states\cite{foot1}.

Is the Majumdar-Gosh or dimer state in the universality class
generic to one-dimensional spin-\half liquids, and hence a
useful paradigm to understand, say, the nearest-neighbor Heisenberg
chain?  The answer is clearly no, as the dimer states (\ref{e:psimg})
violate translational symmetry modulo translations by two lattice
spacings, while the generic liquid is invariant.  A useful paradigm for
the generic liquid is provided by the Haldane-Shastry model, which we
will review in section \ref{s:hs}

Nonetheless, the dimer chain shares some important properties of this
generic liquid.  First, the spinon excitations\cite{foot3}---here domain 
walls between ``even'' and ``odd'' ground states---are free (rather than
confined).  Second, there are (modulo the overall two-fold degeneracy)
only $M+1$ orbitals available for the spinons if $2M$ spins are
condensed into dimers or valence bond singlets.  This is to say, if
there are only a few spinons in a long chain, the number of orbitals
available to them is roughly half the number of sites.  This can
easily be seen graphically:
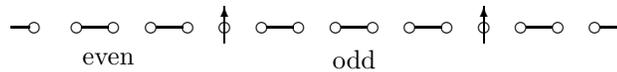
\begin{center}
\begin{picture}(222,26)(-16,-18)
\put(-12,0){\makebox(0,0)[r]{\rule{7.pt}{ 0.8pt}}}
\put(13,0){\makebox(0,0){\rule{10.pt}{ 0.8pt}}}
\put(41,0){\makebox(0,0){\rule{10.pt}{ 0.8pt}}}
\put(83,0){\makebox(0,0){\rule{10.pt}{ 0.8pt}}}
\put(111,0){\makebox(0,0){\rule{10.pt}{ 0.8pt}}}
\put(139,0){\makebox(0,0){\rule{10.pt}{ 0.8pt}}}
\put(181,0){\makebox(0,0){\rule{10.pt}{ 0.8pt}}}
\put(204,0){\makebox(0,0)[l]{\rule{7.pt}{ 0.8pt}}}
\put(-10,0){\circle{4}}
\put(6,0){\circle{4}}
\put(20,0){\circle{4}}
\put(34,0){\circle{4}}
\put(48,0){\circle{4}}
\put(62,0){\circle{4}}
\put(76,0){\circle{4}}
\put(90,0){\circle{4}}
\put(104,0){\circle{4}}
\put(118,0){\circle{4}}
\put(132,0){\circle{4}}
\put(146,0){\circle{4}}
\put(160,0){\circle{4}}
\put(174,0){\circle{4}}
\put(188,0){\circle{4}}
\put(202,0){\circle{4}}
\put(62,1){\makebox(0,0){\vector(0,1){14}}}
\put(160,1){\makebox(0,0){\vector(0,1){14}}}
\put(18,-12){\makebox(0,0){\small even}}
\put(111,-12){\makebox(0,0){\small odd}}
\end{picture}
\end{center}
If we start with an even ground state on the left, the spinon to its
right must occupy an even lattice site and vice versa.  The resulting
state counting is precisely what we will find in the Haldane-Shastry
model, where it is directly linked to the half-fermi statistics of the
spinons\cite{excl}.

The dimer chain is further meaningful as a piece of a general paradigm.  The
two degenerate dimer states (\ref{e:psimg}) can be combined into an
$s=1$ chain, the AKLT chain, which serves as a generic paradigm for
$s=1$ chains which exhibit the Haldane gap\cite{hgap,affl}.  
To formulate the AKLT model,
we will avail ourselves of Schwinger bosons, which we review first.

\section{SCHWINGER BOSONS}
\label{s:schw}

Schwinger bosons\cite{schw,auer} constitute a way to formulate spin-$s$ 
representations of an SU(2) algebra\cite{foot2}.
The spin operators
\begin{equation}
\begin{array}{rcccl}
S^x + iS^y &=& S^+ &=& a^\dagger b \\\rule{0pt}{12pt} 
S^x - iS^y &=& S^- &=& b^\dagger a \\\rule{0pt}{12pt}
&& S^z  &=& \frac{1}{2}(a^\dagger a - b^\dagger b) 
\label{e:schw}
\end{array}
\end{equation}
are given in terms of boson creation and annihilation
operators which obey the usual commutation relations
\begin{equation}
\begin{array}{c}
\comm{a}{a^\dagger}=\comm{b}{b^\dagger}=1 \\
\comm{a}{b}=\comm{a}{b^\dagger}
=\comm{a^\dagger}{b}=\comm{a^\dagger}{b^\dagger}=0\rule{0pt}{16pt}.
\end{array}
\label{e:schwb}
\end{equation}
It is readily verified with (\ref{e:schwb}) that
\begin{equation}
\comm{S^i}{S^j}=i\epsilon^{ijk}S^k 
\quad\hbox{where}\quad i,j,k=x,y,\hbox{or}\ z. 
\label{e:sutwoal}
\end{equation}
The spin quantum number $s$ is given by half the number of bosons,  
\begin{equation}
2s=a^\dagger a + b^\dagger b,
\label{e:schwt}
\end{equation}
and the usual spin states (simultaneous eigenstates of $\boldsymbol{S}^2$
and $S^z$) are given by
\begin{equation}
\ket{s,m} = \frac{(a^\dagger)^{s+m}}{\sqrt{(s+m)!}} \frac{(b^\dagger)^{s-m}}
{\sqrt{(s-m)!}} \vac.
\label{e:schs}
\end{equation}
In particular, the spin-\half states are given by
\begin{equation}
\ket{\up}=a^\dagger \vac =c_{\up}^\dagger \vac \qquad 
\ket{\dw}=b^\dagger \vac =c_{\dw}^\dagger \vac ,
\label{e:schwh}
\end{equation}
\ie $a^\dagger$ and $b^\dagger$ act just like the fermion creation
operators $c^\dagger_\up$ and $c^\dagger_\dw$ in this case.  The
difference shows up only when two (or more) creation operators act on
the same site or orbital.  The fermion operators create an
antisymmetric or singlet configuration (in accordance with the Pauli
principle), 
\begin{equation}
\ket{0,0} = c_{\up}^\dagger c_{\dw}^\dagger \vac ,
\end{equation}
while the Schwinger bosons create a totally symmetric or
triplet (or higher spin if we create more than two bosons) configuration,
\begin{eqnarray}
\ket{1,1} &=& \textstyle{\frac{1}{\sqrt{2}}} (a^\dagger)^{2}\vac \nonumber\\[1mm]
\ket{1,0} &=& a^\dagger b^\dagger \vac \\[1mm]
\ket{1,-1} &=& \textstyle{\frac{1}{\sqrt{2}}} (b^\dagger)^{2}\vac .\nonumber
\end{eqnarray}
Accordingly, may rewrite the Majumdar-Gosh states as
\begin{equation}
\big|\psi^{\scriptscriptstyle\textrm{MG}}_{\textrm{even}\rule{0pt}{5pt}\atop 
\textrm{(odd)}} \big\rangle \, =
\underbrace{\displaystyle
\prod_{i\ \textrm{even}\atop (i\ \textrm{odd})}  
(a^\dagger_{i} b^\dagger_{i+1} - b^\dagger_{i} a^\dagger_{i+1})
}_{\displaystyle
\equiv\Psi^{\scriptscriptstyle\textrm{MG}}_{\textrm{even}\rule{0pt}{5pt}\atop 
\textrm{(odd)}} \big[a^\dagger,b^\dagger\big]
} \,\vac
\label{e:mgschw}
\end{equation}
This formulation readily suggests a generalization to higher spin.

\section{THE AKLT CHAIN}
\label{s:aklt}

Affleck, Kennedy, Lieb and Tasaki\cite{aklt} noticed that the 
valance bond solid state 
\begin{eqnarray}
\ket{\psi^{\scriptscriptstyle\textrm{AKLT}}} \, &=&
\prod_i\, (a^\dagger_{i} b^\dagger_{i+1} - b^\dagger_{i} a^\dagger_{i+1})
\,\vac =\nonumber\\
&=&
\Psi^{\scriptscriptstyle\textrm{MG}}_{\scriptscriptstyle\textrm{even}}
\big[a^\dagger,b^\dagger\big] \,\cdot\, 
\Psi^{\scriptscriptstyle\textrm{MG}}_{\scriptscriptstyle\textrm{odd}}
\big[a^\dagger,b^\dagger\big] \,\vac =\nonumber\\ 
\label{e:psiaklt}
&=&
\ket{\hbox{\begin{picture}(132,8)(-4,-3)
\put(48,0){\circle{10}}
\put(48,-5){\makebox(0,0)[t]{\rule{.3pt}{10pt}}}
\put(4,-22){\makebox(0,0)[l]{\small projections onto spin $s=1$}}
\put(13,1.5){\makebox(0,0){\rule{14.pt}{ 0.8pt}}}
\put(41,1.5){\makebox(0,0){\rule{14.pt}{ 0.8pt}}}
\put(69,1.5){\makebox(0,0){\rule{14.pt}{ 0.8pt}}}
\put(97,1.5){\makebox(0,0){\rule{14.pt}{ 0.8pt}}}
\put(118,1.5){\makebox(0,0)[l]{\rule{5.pt}{ 0.8pt}}}
\put(6,-1.5){\makebox(0,0)[r]{\rule{5.pt}{ 0.8pt}}}
\put(27,-1.5){\makebox(0,0){\rule{14.pt}{ 0.8pt}}}
\put(55,-1.5){\makebox(0,0){\rule{14.pt}{ 0.8pt}}}
\put(83,-1.5){\makebox(0,0){\rule{14.pt}{ 0.8pt}}}
\put(111,-1.5){\makebox(0,0){\rule{14.pt}{ 0.8pt}}}
\end{picture}}}\rule{0pt}{18pt}\\ \nonumber \rule{0pt}{15pt}
\end{eqnarray}
is the exact zero energy ground state of the spin 1 extended 
Heisenberg Hamiltonian
\begin{equation}
H^{\scriptscriptstyle\textrm{AKLT}} = 
\sum_i \left(\boldsymbol{S}_i \boldsymbol{S}_{i+1} +
\frac{1}{3}\left(\boldsymbol{S}_i \boldsymbol{S}_{i+1}\right)^2 
+\frac{2}{3}\right) 
\label{e:haklt}
\end{equation}
with periodic boundary conditions.
Each term in the sum (\ref{e:haklt}) projects onto the subspace in which 
the total spin of a pair of neighboring sites is $s=2$.  The Hamiltonian
(\ref{e:haklt}) thereby lifts all states except (\ref{e:psiaklt}) to
higher energies\cite{proof}.  The valance bond solid state
(\ref{e:psiaklt}) is a generic paradigm as it shares all the
symmetries, but in particular the Haldane spin gap\cite{hgap,affl}, 
of the spin-1
Heisenberg chain.  It even offers a particularly simple understanding
of this gap, or of the linear confinement force between spinons 
responsible for it, as illustrated by a cartoon:
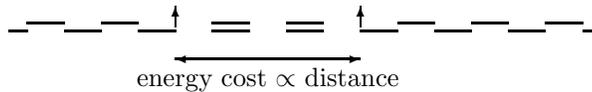
\begin{center}
\begin{picture}(230,40)(-32,-26)
\put(-15,1.5){\makebox(0,0){\rule{14.pt}{ 0.8pt}}}
\put(13,1.5){\makebox(0,0){\rule{14.pt}{ 0.8pt}}}
\put(55,1.5){\makebox(0,0){\rule{14.pt}{ 0.8pt}}}
\put(83,1.5){\makebox(0,0){\rule{14.pt}{ 0.8pt}}}
\put(125,1.5){\makebox(0,0){\rule{14.pt}{ 0.8pt}}}
\put(153,1.5){\makebox(0,0){\rule{14.pt}{ 0.8pt}}}
\put(181,1.5){\makebox(0,0){\rule{14.pt}{ 0.8pt}}}
\put(-22,-1.5){\makebox(0,0)[r]{\rule{7.pt}{ 0.8pt}}}
\put(-1,-1.5){\makebox(0,0){\rule{14.pt}{ 0.8pt}}}
\put(27,-1.5){\makebox(0,0){\rule{14.pt}{ 0.8pt}}}
\put(55,-1.5){\makebox(0,0){\rule{14.pt}{ 0.8pt}}}
\put(83,-1.5){\makebox(0,0){\rule{14.pt}{ 0.8pt}}}
\put(111,-1.5){\makebox(0,0){\rule{14.pt}{ 0.8pt}}}
\put(139,-1.5){\makebox(0,0){\rule{14.pt}{ 0.8pt}}}
\put(167,-1.5){\makebox(0,0){\rule{14.pt}{ 0.8pt}}}
\put(188,-1.5){\makebox(0,0)[l]{\rule{7.pt}{ 0.8pt}}}
%
\put(34,0){{\vector(0,1){8}}}
\put(104,0){{\vector(0,1){8}}}
\put(69,-12){\vector(-1,0){35}}
\put(69,-12){\vector(1,0){35}}
\put(69,-20){\makebox(0,0){\small energy cost $\propto$ distance}}
\end{picture}
\end{center}

We have reviewed the AKLT model here as the simplest and the only
generally known application of the general projection principle
described in the introduction.  The Haldane gap illustrates that not
only the symmetries, but also the physical properties, and in
particular the low energy properties, may be very different for the
combined spin-$ms$ states as for the original spin-$s$ state.

\section{THE HALDANE-SHASTRY MODEL}
\label{s:hs}

The Haldane-Shastry model\cite{hs,morehs,mimo} is the most important 
paradigm for a generic spin-\half liquid.  Consider a spin-\half chain with
periodic boundary conditions and an even number of sites $N$ on a unit
circle embedded in the complex plane:
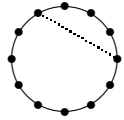
\begin{center}
\begin{picture}(320,70)(-40,-35)
\put(0,0){\circle{100}}
\put( 20.0,   .0){\circle*{3}}
\put( 17.3, 10.0){\circle*{3}}
\put( 10.0, 17.3){\circle*{3}}
\put(   .0, 20.0){\circle*{3}}
\put(-10.0, 17.3){\circle*{3}}
\put(-17.3, 10.0){\circle*{3}}
\put(-20.0,   .0){\circle*{3}}
\put(-17.3,-10.0){\circle*{3}}
\put(-10.0,-17.3){\circle*{3}}
\put(   .0,-20.0){\circle*{3}}
\put( 10.0,-17.3){\circle*{3}}
\put( 17.3,-10.0){\circle*{3}}
\qbezier[20]( 20.0,   .0)(  5.0,8.65)(-10.0, 17.3)
\put(50,12){\makebox(0,0)[l]
{$N$\ sites with spin \half on unit circle: 
}}
\put(50,-12){\makebox(0,0)[l]
{$\displaystyle \eta_\alpha=e^{i\frac{2\pi}{N}\alpha }$
\ \ with\ $\alpha = 0,1,\ldots ,N\! -\! 1$}}
\end{picture}
\end{center}
The ${1}/{r^2}$-Hamiltonian
\begin{equation}
H^{\scriptscriptstyle\textrm{HS}} = \left(\frac{2\pi}{N}\right)^2\,
\sum_{\alpha <\beta}\,
\frac{\boldsymbol{S}_\alpha \boldsymbol{S}_\beta 
}{\left|\eta_\alpha-\eta_\beta \right|^2}\,,
\label{e:hhs}
\end{equation}
where $\left|\eta_\alpha-\eta_\beta \right|$ is the chord distance between
the sites $\alpha$ and $\beta$, has the exact ground state 
\begin{equation}
\ket{\psi^{\scriptscriptstyle\textrm{HS}}_{\scriptscriptstyle 0}}\,=
\sum_{\{z_1,z_2,\ldots ,z_M\}}\, 
\psi^{\scriptscriptstyle\textrm{HS}}_{\scriptscriptstyle 0} 
(z_1,z_2,\ldots ,z_M)\,\,S^+_{z_1}\cdot\ldots\cdot S^+_{z_M}\,\, 
\big|\underbrace{\dw\dw\ldots\ldots\dw
}_{\textrm{all\ } N \textrm{\ spins\ } \dw}
\big\rangle
\label{e:keths}
\end{equation}
where the sum extends over all possible ways to distribute the 
$M=\frac{N}{2}$ $\up$-spin coordinates $z_i$ and
\begin{equation}
\psi^{\scriptscriptstyle\textrm{HS}}_{\scriptscriptstyle 0}
(z_1,z_2,\ldots ,z_M) = 
\prod_{j<k}^M\,(z_j-z_k)^2\,\prod_{j=1}^M\,z_j\,. 
\label{e:psihs}
\end{equation}
This state 
is real, a spin singlet, and has ground state energy
$-\frac{\pi^2}{24}\left(N-\frac{5}{N}\right)$.

The proof of solution is rather lengthy\cite{hs,mimo}.  We will 
content ourselves here to showing that (\ref{e:psihs}) is real,
and that it is a spin singlet.  As for the former, we use 
\begin{equation}
(z_j-z_k)^2\,=- z_j z_k\, |z_j-z_k|^2\,,
\end{equation}
to write
\begin{eqnarray}
\psi^{\scriptscriptstyle\textrm{HS}}_{\scriptscriptstyle 0}
(z_1,z_2,\ldots ,z_M) &=&
\pm\,\prod_{j<k}^M\,|z_j-z_k|^2\;
\prod_{j<k}^M\,z_j z_k\;\prod_{j=1}^M\,z_j\,=\nonumber\\
&=&\pm\,\prod_{j<k}^M\,|z_j-z_k|^2\;\prod_{j=1}^M\,G(z_j)
\label{e:psihsreal}
\end{eqnarray}
where 
\begin{equation}
G(\eta_\alpha)=(\eta_\alpha)^\frac{N}{2}=
\left\{\begin{array}{ll} 
+1 &\quad \alpha\ \textrm{even}\\
-1 &\quad \alpha\ \textrm{odd}.
\end{array}\right.
\label{e:psihsg}
\end{equation}
The gauge factor $G(z_j)$ effects that the Marshall sign criteria\cite{mars} 
is fulfilled.  

We now proof that (\ref{e:psihs}) is a singlet.  Since
$S^z_{\textrm{tot}}\,
\ket{\psi^{\scriptscriptstyle\textrm{HS}}_{\scriptscriptstyle 0}}\,
=0$, it suffices to show that
$\ket{\psi^{\scriptscriptstyle\textrm{HS}}_{\scriptscriptstyle 0}}$ 
is annihilated by $S^-_{\textrm{tot}}$:
\begin{eqnarray}
S^-_{\textrm{tot}}
\ket{\psi^{\scriptscriptstyle\textrm{HS}}_{\scriptscriptstyle 0}}\!&=&
\sum_{\alpha=1}^{N} S^-_\alpha 
\sum_{\{z_1,\ldots z_M\}}
\psi^{\scriptscriptstyle\textrm{HS}}_{\scriptscriptstyle 0} 
(z_1,z_2,\ldots z_M)\, S^+_{z_1}\ldots S^+_{z_M}\ket{\dw\ldots\dw}= 
\quad\nonumber\\ 
&=&\sum_{\{z_2,\ldots,z_M\}}\, 
\underbrace{\sum_{\alpha=1}^{N}\,
\psi^{\scriptscriptstyle\textrm{HS}}_{\scriptscriptstyle 0} 
(\eta_\alpha,z_2,\ldots,z_M)}_{=0} 
\,S^+_{z_2}\ldots S^+_{z_M}\ket{\dw\ldots\dw}
\end{eqnarray}
since $\psi^{\scriptscriptstyle\textrm{HS}}_{\scriptscriptstyle 0} 
(\eta_\alpha,z_2,\ldots,z_M)$ contains powers 
$\eta_\alpha^1, \eta_\alpha^2,\ldots , \eta_\alpha^{N-1}$ and
\begin{equation}
\sum_{\alpha=1}^{N} \eta_\alpha^n = 
N\delta_{n,0}\quad \textrm{mod}\ N.
\end{equation}

The elementary excitations for this model are deconfined and only
weak\-ly interacting spinon excitations, which carry spin-\half and no
charge.  They constitute an instance of fractional quantization,
similar (both conceptually and mathematically) to the fractional
quantization of charge in the fractional quantum Hall
effect\cite{fqhe}.  Their fractional quantum number is the spin, which
takes the value \half in a Hilbert space (\ref{e:keths}) made out of
spin flips $S^+$, which carry spin 1.

To write down the wave for a $\dw$ spin spinon localized at site 
$\eta_\alpha$, consider a chain with an odd number of sites $N$ and
let $M=\frac{N-1}{2}$ be the number of $\up$ or $\dw$ spins condensed
in the uniform liquid.  The spinon is then given by
\begin{equation}
\psi^{\scriptscriptstyle\textrm{HS}}_{\alpha\dw}
(z_1,z_2,\ldots ,z_M) =
\prod_{j=1}^M\,(\eta_\alpha -z_j)\,
\prod_{j<k}^M\,(z_j-z_k)^2\,\prod_{j=1}^M\,z_j\, ,
\label{e:psihssp}
\end{equation}
which we again understand substituted into (\ref{e:keths}).  
It is easy to verify that
$S^z_{\textrm{tot}} \psi^{\scriptscriptstyle\textrm{HS}}_{\alpha\dw} = 
-\frac{1}{2} \psi^{\scriptscriptstyle\textrm{HS}}_{\alpha\dw}$ and that
$S^-_{\textrm{tot}} \psi^{\scriptscriptstyle\textrm{HS}}_{\alpha\dw} = 0$,
which shows that the spinon transforms as a spinor under rotations.  

The localized spinon (\ref{e:psihssp}) is not an eigenstate of the
Haldane-Shastry Hamiltonian (\ref{e:hhs}).  To obtain exact
eigenstates, we construct momentum eigenstates according to
\begin{equation}
\psi^{\scriptscriptstyle\textrm{HS}}_{m\dw}
(z_1,z_2,\ldots ,z_M) =
\,\sum_{\alpha=1}^{N} \overline{\eta}_\alpha^m\,
\psi^{\scriptscriptstyle\textrm{HS}}_{\alpha\dw}
(z_1,z_2,\ldots ,z_M). 
\label{e:psihsmo}
\end{equation}
where the integer $m$ corresponds to a momentum quantum number.
Since 
$\psi^{\scriptscriptstyle\textrm{HS}}_{\alpha\dw}
(z_1,z_2,\ldots ,z_M)$ contains only powers 
$\eta_\alpha^0, \eta_\alpha^1,\ldots , \eta_\alpha^M$ and 
\begin{equation}
\sum_{\alpha=1}^{N} \overline{\eta}_\alpha^m \eta_\alpha^n = \delta_{mn}
\quad \textrm{mod}\ N,
\label{e:hsdelta}
\end{equation}
$\psi^{\scriptscriptstyle\textrm{HS}}_{m\dw} (z_1,z_2,\ldots ,z_M)$
will vanish unless $m=0,1,\ldots ,M$.  There are only roughly half
as many spinon orbitals as sites.  Spinons on neighboring sites 
hence cannot be orthogonal.

To make a correspondence between $m$ and the usual wave number 
$q$, we translate (\ref{e:psihsmo}) counterclockwise by
one lattice spacing around the unit circle
\begin{equation}
\boldsymbol{T}\, \ket{\psi^{\scriptscriptstyle\textrm{HS}}_{m\dw}} 
= e^{iq} \ket{\psi^{\scriptscriptstyle\textrm{HS}}_{m\dw}},
\label{e:hst}
\end{equation}
and find
\begin{equation}
q=\pi M + \pi\frac{M-2m}{2M+1}.
\label{e:hsq}
\end{equation}
The spinon dispersion is given by 
\begin{equation}
E_q=\frac{1}{2}\left[\big(\frac{\pi}{2}\big)^2-q^2\right]
\quad \textrm{mod}\ \pi,
\label{e:hseq}
\end{equation}
as depicted below.
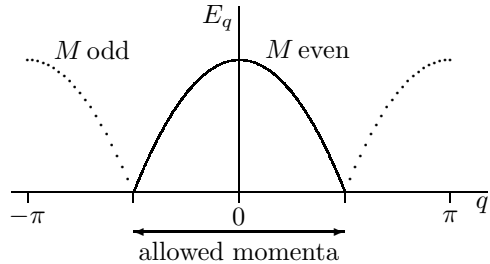
\begin{center}
\begin{picture}(200,110)(-100,-30)
\qbezier[2000](-40,0)(0,100)(40,0)
\put(4,0){\makebox(0,0){\rule{180.pt}{0.3pt}}}
\put(0,0){\makebox(0,0)[b]{\rule{0.3pt}{70.pt}}}
\put(-80,0){\makebox(0,0)[t]{\rule{0.3pt}{4pt}}}
\put(-80,-9){\makebox(0,0){\small $-\pi$}}
\put(-40,0){\makebox(0,0)[t]{\rule{0.3pt}{4pt}}}
\put(0,0){\makebox(0,0)[t]{\rule{0.3pt}{4pt}}}
\put(0,-9){\makebox(0,0){\small 0}}
\put(40,0){\makebox(0,0)[t]{\rule{0.3pt}{4pt}}}
\put(80,0){\makebox(0,0)[t]{\rule{0.3pt}{4pt}}}
\put(80,-9){\makebox(0,0){\small $\pi$}}
\put(92,-5){\makebox(0,0){\small $q$}}
\put(-8,66){\makebox(0,0){\small $E_q$}}
\put(0,-15){{\vector(-1,0){40}}}
\put(0,-15){{\vector(1,0){40}}}
\put(0,-22){\makebox(0,0){\small allowed momenta}}
\put(10,54){\makebox(0,0)[l]{\small $M\!$ even}}
\put(-70,54){\makebox(0,0)[l]{\small $M\!$ odd}}
\put(-80.0, 50.0){\circle*{1}}
\put(-78.0, 49.9){\circle*{1}}
\put(-76.0, 49.5){\circle*{1}}
\put(-74.0, 48.9){\circle*{1}}
\put(-72.0, 48.0){\circle*{1}}
\put(-70.0, 46.9){\circle*{1}}
\put(-68.0, 45.5){\circle*{1}}
\put(-66.0, 43.9){\circle*{1}}
\put(-64.0, 42.0){\circle*{1}}
\put(-62.0, 39.9){\circle*{1}}
\put(-60.0, 37.5){\circle*{1}}
\put(-58.0, 34.9){\circle*{1}}
\put(-56.0, 32.0){\circle*{1}}
\put(-54.0, 28.9){\circle*{1}}
\put(-52.0, 25.5){\circle*{1}}
\put(-50.0, 21.9){\circle*{1}}
\put(-48.0, 18.0){\circle*{1}}
\put(-46.0, 13.9){\circle*{1}}
\put(-44.0,  9.5){\circle*{1}}
\put(-42.0,  4.9){\circle*{1}}
%
\put( 80.0, 50.0){\circle*{1}}
\put( 78.0, 49.9){\circle*{1}}
\put( 76.0, 49.5){\circle*{1}}
\put( 74.0, 48.9){\circle*{1}}
\put( 72.0, 48.0){\circle*{1}}
\put( 70.0, 46.9){\circle*{1}}
\put( 68.0, 45.5){\circle*{1}}
\put( 66.0, 43.9){\circle*{1}}
\put( 64.0, 42.0){\circle*{1}}
\put( 62.0, 39.9){\circle*{1}}
\put( 60.0, 37.5){\circle*{1}}
\put( 58.0, 34.9){\circle*{1}}
\put( 56.0, 32.0){\circle*{1}}
\put( 54.0, 28.9){\circle*{1}}
\put( 52.0, 25.5){\circle*{1}}
\put( 50.0, 21.9){\circle*{1}}
\put( 48.0, 18.0){\circle*{1}}
\put( 46.0, 13.9){\circle*{1}}
\put( 44.0,  9.5){\circle*{1}}
\put( 42.0,  4.9){\circle*{1}}
%
\end{picture}
\end{center}
The construction (\ref{e:psihssp},\ref{e:psihsmo}) can be generalized 
to many spinon states, which weakly attract each other.  Then the 
individual spinon momenta labeled by $m_l$ ($l=1,2,\ldots ,L$, where
$L$ is the number of spinons) are no longer good quantum numbers,
as they have a channel to decay into. 
The construction can, however, still be used to count the number of 
available orbitals for spinons, which leads us to the subject of 
fractional statistics.

\section{FRACTIONAL STATISTICS IN SPIN LIQUIDS}
\label{s:frac}

Consider a Haldane-Shastry chain with an even number of sites $N$ and
an even number of spinons $L$.  According to
(\ref{e:psihssp}),(\ref{e:psihsmo}) and (\ref{e:hsdelta}), the number
of orbitals available to each spinon is given by $M+1$, where
$M=\frac{N-L}{2}$ is the number of $\up$ or $\dw$ spins in the
remaining uniform liquid.
(In this representation, the spinon wave functions are symmetric; two
or more spinons can have the same value for $m$.)  The creation of
{\em two} spinons reduces the number of available orbitals hence by
{\em one}.  They obey half-fermi statistics in the sense of
Haldane's exclusion principle\cite{excl}.  (For fermions, the creation
of two particles would decrease the number of available orbitals by
two, while this number would not change for bosons.)

It is very instructive to use this fractional exclusion principle
to count the dimension of the Hilbert space spanned by the ground
state plus all possible spinon configurations\cite{excl}.  In general, 
the number of ways to place $x$ bosons (as the spinon wave functions 
are symmetric---the spinons are half-fermions formulated in a bosonic
representation) into $y$ orbitals is given by
\begin{equation}
\frac{(x+y-1)!}{x!\,\, (y-1)!}\,.
\end{equation}
For a configuration with $L$ spinons, we have $x=L$ and 
$y=2 (M+1)=N-L+2$ orbitals, where the factor 2 stems from the internal
spin degeneracy of each spinon orbital.  The total number of states 
is 
\begin{equation}
\sum_{\scriptstyle L=0 \atop \scriptstyle\textrm{even}}^N \,
\frac{(N+1)!}{L!\, (N-L+1)!}\, =\, 2^N
\end{equation}
as it should be for a spin \half system with $N$ sites.

We may interpret the Haldane-Shastry model as a reparameterization 
of a Hilbert space spanned by spin flips (\ref{e:keths}) into a
basis which consists of the Haldane-Shastry ground state
plus all possible spinon excitations.  The reward for such a
reparameterization is that {a highly non-trivial Hamiltonian in the
original basis may be approximately or exactly diagonal in the new
basis.}  This reparameterization is particularly useful as it usually
instructs us directly about the quantum numbers of the elementary
excitations.  In the fractional quantum Hall effect, we have learned
from such a reparameterization---or trial wave function---that the
quasiparticles carry fractional charge.  For the spin \half chain, we
can learn that the elementary excitations are spinons, which carry
spin \half.
 
Of course, it is a condition 
that the Hilbert space dimensions before and after the
reparameterization match.  It is not very difficult to convince oneself
that the fractional state counting, and hence the fractional or half
fermi statistics, is a necessary condition to accomplish this for a
spin \half system with spin \half spinon excitations.  This leads me
to the following hypothesis:
\begin{quote}
{\em Spinons must carry fractional or half-fermion\cite{sutwo} 
statistics, regardless of the
dimension, in order for the states to count up consistently.}
\end{quote}

In one space dimension, fractional statistics can only be defined 
through a fractional exclusion principle, which does not yield
any symmetry violations.  In two space dimensions, however, we can
define statistics alternatively through the Berry's phase 
acquired by the
state as we adiabatically exchange particles by moving them 
counterclockwise around each other\cite{wilc}.  This phase is given 
by $e^{i\theta}$, where $\theta$ is the statistical parameter.
\begin{center}
\begin{picture}(210,50)(-50,-12)
\put(0,0){\oval(40,40)[t]}
\put(0,0){\circle*{3}}
\put(20,0){\circle*{3}}
\put(-20,0){{\vector(0,-1){2}}}
\put(60,6){\makebox(0,0)[l]
{$\displaystyle \ket{\psi}\,\rightarrow \,e^{i\theta}\,\ket{\psi}$}}
\end{picture}
\end{center}
For $\theta=\pi$ we have fermions,
for $\theta=0$ bosons, and for both $\theta=\frac{\pi}{2}$ and
$\theta=-\frac{\pi}{2}$ half-fermions.  The choice of sign for 
$\theta$ is physically meaningful, as the allowed values for
the relative angular momentum $l$ depend on it:
\begin{equation}
l=\frac{\theta}{\pi} + 2n\quad \textrm{where}\ n\ \textrm{integer.}
\end{equation}
Clearly either choice of sign for half-fermions violates parity (P)
and time reversal (T).  

If we now assume that both definitions of statistics match for spinons
in two dimensional spin liquids, as they do for the quasiparticles in
the fractionally quantized Hall effect, the hypothesis of half fermi
statistics leaves us with only two choices.  The first is that the
spin liquid ground state violates P and T spontaneously and thereby
fixes the sign for the statistical parameter for the spinons.  This
possibility has been advocated by R.B.\ Laughlin.  The chiral spin 
liquid\cite{chiral,wwz,zou}, which we will review in the following section,
provides us with a paradigm for this situation. 

The second possibility, which I would like to advocate, is that
the spinons carry a chirality quantum number, which determines the 
sign of the statistical parameter $\theta$.  Then there is no need
for the ground state to violate any symmetries.  Spinons of different
chiralities map into each other under P and T.  The situation is
somewhat analogous to fermions in one dimension, where we have 
right and left moving excitations, which are P and T conjugates 
of each other.  I will construct the simplest paradigm for such
a liquid, the spin 1 chirality liquid, in section \ref{s:cl}

\section{THE CHIRAL SPIN LIQUID}
\label{s:chiral}

The chiral spin liquid\cite{chiral,wwz,dou,zou} may be viewed as a
brute-force generalization of the Haldane-Shastry wave function
to two space dimensions\cite{unfair}.  Consider a
periodic one-dimensional lattice on the real axis of a complex plane,
with lattice points at integer values:
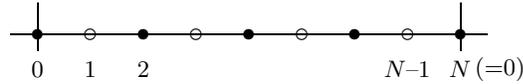
\begin{center}
\begin{picture}(200,40)(-100,-20)
\put(0,0){\makebox(0,0){\rule{180.pt}{ 0.3pt}}}
\put(80,-6){\makebox(0,0)[b]{\rule{ 0.3pt}{18.pt}}}
\put(-80,-6){\makebox(0,0)[b]{\rule{ 0.3pt}{18.pt}}}
\put(80,-13){\makebox(0,0){\footnotesize $N$}}
\put(95,-13){\makebox(0,0){\footnotesize (=$0$)}}
\put(-80,-13){\makebox(0,0){\footnotesize 0}}
\put(60,-13){\makebox(0,0){\footnotesize $N$--$1$}}
\put(-60,-13){\makebox(0,0){\footnotesize 1}}
\put(-40,-13){\makebox(0,0){\footnotesize 2}}
\put(0,0){\circle*{4}}
\put(20,0){\circle{4}}
\put(-20,0){\circle{4}}
\put(40,0){\circle*{4}}
\put(-40,0){\circle*{4}}
\put(60,0){\circle{4}}
\put(-60,0){\circle{4}}
\put(80,0){\circle*{4}}
\put(-80,0){\circle*{4}}
\end{picture}
\end{center}
Filled and empty circles represent even and odd integers, with gauge
factors $G(z)=-1$ and $G(z)=+1$, respectively.
The Haldane-Shastry wave functions (\ref{e:psihsreal}) then becomes
\begin{equation}
\psi^{\scriptscriptstyle\textrm{HS}}_{\scriptscriptstyle 0}
(z_1,\ldots ,z_M)= 
\prod_{j=1}^M\,G(z_j)\;\prod_{j<k}^M\,
\sin\!\left(\frac{\pi}{N} (z_j-z_k)\right)^2,
\label{e:psihssin}
\end{equation}
where we took advantage of the fact that now the coordinates $z_j$ are real.

The chiral spin liquid is obtained by extending the lattice
from a circle to a cylinder, or from a segment of the real axis to a 
strip in the two-dimensional complex plane:  
\begin{center}
\begin{picture}(200,110)(-100,-55)
\put(0,0){\makebox(0,0){\rule{180.pt}{ 0.3pt}}}
\put(80,0){\makebox(0,0){\rule{ 0.3pt}{92.pt}}}
\put(-80,0){\makebox(0,0){\rule{ 0.3pt}{92.pt}}}
\put(88,-7){\makebox(0,0){\footnotesize $L_x$}}
\put(103,-7){\makebox(0,0){\footnotesize (=$0$)}}
\put(-74,-7){\makebox(0,0){\footnotesize 0}}
\put(0,0){\circle*{4}}
\put(20,0){\circle{4}}
\put(-20,0){\circle{4}}
\put(40,0){\circle*{4}}
\put(-40,0){\circle*{4}}
\put(60,0){\circle{4}}
\put(-60,0){\circle{4}}
\put(80,0){\circle*{4}}
\put(-80,0){\circle*{4}}
\put(0,40){\circle*{4}}
\put(20,40){\circle{4}}
\put(-20,40){\circle{4}}
\put(40,40){\circle*{4}}
\put(-40,40){\circle*{4}}
\put(60,40){\circle{4}}
\put(-60,40){\circle{4}}
\put(80,40){\circle*{4}}
\put(-80,40){\circle*{4}}
\put(0,20){\circle{4}}
\put(20,20){\circle{4}}
\put(-20,20){\circle{4}}
\put(40,20){\circle{4}}
\put(-40,20){\circle{4}}
\put(60,20){\circle{4}}
\put(-60,20){\circle{4}}
\put(80,20){\circle{4}}
\put(-80,20){\circle{4}}
\put(0,-40){\circle*{4}}
\put(20,-40){\circle{4}}
\put(-20,-40){\circle{4}}
\put(40,-40){\circle*{4}}
\put(-40,-40){\circle*{4}}
\put(60,-40){\circle{4}}
\put(-60,-40){\circle{4}}
\put(80,-40){\circle*{4}}
\put(-80,-40){\circle*{4}}
\put(0,-20){\circle{4}}
\put(20,-20){\circle{4}}
\put(-20,-20){\circle{4}}
\put(40,-20){\circle{4}}
\put(-40,-20){\circle{4}}
\put(60,-20){\circle{4}}
\put(-60,-20){\circle{4}}
\put(80,-20){\circle{4}}
\put(-80,-20){\circle{4}}
\end{picture}
\end{center}
where $G(z)=(-1)^{(x+1)(y+1)}$ for lattice site $z=x+iy$.  The wave 
function for a chiral spin liquid on this cylinder is given by
(\ref{e:psihssin}) multiplied by an exponential factor,
which effects 
that the density of spin flips is \half for $-L_y/2 < y < L_y/2$,
\ie up to the boundaries of the liquid:
\begin{equation}
\psi^{\scriptscriptstyle\textrm{CSL}}_+
(z_1,\ldots ,z_M)= 
\prod_{j=1}^M\,G(z_j)\;\prod_{j<k}^M\,
\sin\!\left(\frac{\pi}{L_x} (z_j-z_k)\right)^2\;
\prod_{j=1}^M\,e^{-\pi|y_j|^2}.
\label{e:psichisin}
\end{equation}

The chiral spin liquid takes a more familiar form if we consider open 
boundary conditions.  The wave function for a circular droplet of fluid is
simply
\begin{equation}
\psi^{\scriptscriptstyle\textrm{CSL}}_+
(z_1,\ldots ,z_M)= 
\prod_{j=1}^M\,G(z_j)\;\prod_{j<k}^M\,(z_j-z_k)^2\;
\prod_{j=1}^M\,e^{-\frac{\pi}{2}|z_j|^2}.
\label{e:psichi}
\end{equation}
Note that the exponential in (\ref{e:psichisin}) or (\ref{e:psichi}) 
corresponds to a (fictitious) magnetic field of strength $2\pi$/plaquet.  
Apart from the gauge factors $G(z_j)$, (\ref{e:psichi}) is formally
equivalent to the wave function for a fractionally quantized Hall liquid
for bosons at filling fraction $\nu =\frac{1}{2}$, which corresponds to
half a particle (or spin flip) per plaquet.  
$\psi^{\scriptscriptstyle\textrm{CSL}}_+$ is a spin singlet.

The chiral spin liquid is in general complex as the $z$'s are complex.
There are two ``degenerate'' states,
$\psi^{\scriptscriptstyle\textrm{CSL}}_+$ and its complex conjugate
$\psi^{\scriptscriptstyle\textrm{CSL}}_-$, which are P and T
conjugates of each other.  (I have written ``degenerate'' in quotation
marks as no-one has ever been able to construct a local parent
Hamiltonian for this liquid.)  Since the chiral spin liquid violates
the discrete symmetries P and T, it is not in the same universality
class as the generic disordered spin \half antiferromagnet in two space
dimensions, which I presume to be stabilized for some lattice type or
upon doping.

It is rather easy to formulate spinon excitations for the chiral spin 
liquid.  In analogy to both the quasiholes in fractionally quantized
Hall liquids and the spinons (\ref{e:psihssp}) for the Haldane-Shastry 
model, we write the wave function for a spinon localized at $\eta $ 
\begin{equation}
\psi^{\scriptscriptstyle\textrm{CSL}}_{+\eta\dw}
(z_1,\ldots,z_M) =
\prod_{j=1}^M\,(\eta_\alpha -z_j)\;
\prod_{j=1}^M\,G(z_j)\;\prod_{j<k}^M\,(z_j-z_k)^2\;
\prod_{j=1}^M\,e^{-\frac{\pi}{2}|z_j|^2}.
\label{e:psichisp}
\end{equation}
The spinons obey half-fermi statistics, both in the sense of Haldane's
exclusion principle as well as in the sense of the
Berry's phases encountered by the state as we exchange particles
by winding them counterclockwise 
around each other.  We can evaluate this phase using the adiabatic
transport argument of Arovas, Schrieffer, and Wilczek\cite{arov}. 
We obtain $\pi/2$ for (\ref{e:psichisp}) and $-\pi/2$ for its
complex conjugate. 

The chiral spin liquid is a wonderful paradigm, with only one 
imperfection: the violation of P and T.  There is no indication that 
these symmetries are spontaneously broken in any two-dimensional 
spin system.

Like in the case of the Majumdar-Gosh state discussed in section 
\ref{s:mg}, the chiral spin liquid can be used to construct
a symmetry invariant spin liquid with $s=1$, the spin 1 chirality liquid.
For this construction, it is propitious to
formulate the chiral spin liquid 
in a basis which constitutes of electron creation 
operators rather than spin flips.  
To begin with, we use $M=N/2$ 
to rewrite (\ref{e:psichi}) as
\begin{equation}
\psi^{\scriptscriptstyle\textrm{CSL}}_+ (z_1,\ldots,z_M) =
\prod_{j=1}^M\bigg( G(z_j)\,e^{-\frac{M\pi}{N}|z_j|^2}
\prod_{\scriptstyle k=1 \atop\scriptstyle(k\ne j)}^M\,(z_j-z_k)\,
\bigg). 
\label{e:psichi2}
\end{equation}
Let $w_l$ with $l=1,2,\ldots,N-M$ be the lattice sites not occupied by
the $z_j$'s.  Then the octopus theorem\cite{chiral}
\begin{equation}
G(z_j)\,e^{-\frac{\pi}{2}|z_j|^2}\!\!
\prod_{\scriptstyle\alpha =1 \atop\scriptstyle (\eta_\alpha\ne z_j)}^N\!\!
(z_j- \eta_\alpha) \,=\,
\prod_{\alpha =1}^N e^{+\frac{\pi}{2N}|\eta_\alpha |^2}\cdot
\textrm{const.},
\label{e:oct}
\end{equation}
where the first product runs over all lattice sites except $z_j$, implies
\begin{equation}
\psi^{\scriptscriptstyle\textrm{CSL}}_+ (z_1,\ldots,z_M) =
\prod_{j=1}^M\left(\, \prod_{l=1}^{M}\frac{1}{z_j-w_l}\, \right)\;
\prod_{j=1}^M e^{+\frac{(N-M)\pi}{2N}|z_j|^2}\,
\prod_{l=1}^{M} e^{+\frac{M\pi}{2N}|w_l|^2}.
\label{e:psichi3}
\end{equation}
Let
\begin{equation}
\mathcal{S}[z,w]\,\equiv\,\bra{\,0\,}
c_{z_1\dw}
\ldots c_{z_M\dw}\,
c_{w_1\dw}
\ldots c_{w_{N-M}\dw}
\big|\underbrace{\dw\dw\ldots\ldots\dw
}_{\textrm{all\ } N \textrm{\ spins\ } \dw}
\big\rangle,
\label{e:sign}
\end{equation}
be the sign associated with ordering the $z$'s and $w$'s according to 
their lattice positions.  Then we can use
\begin{equation}
\begin{array}{r}\displaystyle
\prod_{j<k}^M(z_j-z_k)\;
\prod_{j=1}^M\prod_{l=1}^{N-M}(z_j-w_l)\;
\prod_{l<m}^{N-M}(w_l-w_m)\,\cdot\rule{50pt}{0pt} 
\\ \rule{0pt}{27pt}\displaystyle\cdot\,
\prod_{j=1}^M e^{-\frac{\pi}{2}|z_j|^2}\,
\prod_{l=1}^{N-M} e^{-\frac{\pi}{2}|w_l|^2}\,= \,\mathcal{S}[z,w]
\cdot\textrm{const.}
\end{array}
\label{e:allsites1}
\end{equation}
to rewrite (\ref{e:psichi3}) as
\begin{equation}
\psi^{\scriptscriptstyle\textrm{CSL}}_+ (z_1,\ldots,z_M) \,=\,
\mathcal{S}[z,w]\;\phi (z_1,\ldots,z_M)\;\phi (w_1,\ldots,w_{N-M}) 
\label{e:psichi4}
\end{equation}
where 
\begin{equation}
\phi (z_1,\ldots,z_M)\,=
\,\prod_{j<k}^M(z_j-z_k)\,\prod_{j=1}^M e^{-\frac{M\pi}{2N}|z_j|^2}
\label{e:ll}
\end{equation}
is simply the wave function for a filled Landau level in a (fictitious) 
magnetic field with flux $\frac{2\pi M}{N}$/plaquet.  If we rewrite 
(\ref{e:psichi}) in terms of (\ref{e:ll}) and compare it to 
(\ref{e:psichi4}), we obtain the lattice particle-hole symmetry
\begin{equation}
\prod_{j=1}^M G(z_j)\ \phi (z_1,\ldots,z_M)\,=\,
\mathcal{S}[z,w]\,\phi (w_1,\ldots,w_{N-M})\cdot\textrm{const.},
\label{e:zw}
\end{equation}
which holds for any $M$.  The chiral spin liquid ground state, where
$M=N/2$, is according to (\ref{e:psichi4}) simply given by
\begin{equation}
\begin{array}{r}\displaystyle
\ket{\psi^{\scriptscriptstyle\textrm{CSL}}_+}
=\sum_{\scriptstyle\{z_1,\ldots ,z_M; w_1,\ldots,w_{M} \}}\,
\phi (z_1,\ldots,z_M)\,\phi (w_1,\ldots,w_{M})\,\cdot\rule{15pt}{0pt} 
\\ \rule{0pt}{0pt}\displaystyle
\cdot\,c^\dagger_{z_1\up}\ldots c^\dagger_{z_M\up}\,
       c^\dagger_{w_1\dw}\ldots c^\dagger_{w_{M}\dw}\vac ,
\end{array}
\label{e:csl}
\end{equation}
where the sum extends over all possible ways to distribute the
coordinates $z_j$ and $w_l$ on mutually distinct lattice sites.  
It is often convenient to write (\ref{e:csl}) as
\begin{equation}
\ket{\psi^{\scriptscriptstyle\textrm{CSL}}_+} \,=\, 
\textrm{P}_{\scriptscriptstyle\textrm{GW}} 
\ket{\psi^N_{\scriptscriptstyle\textrm{SD}}}
\label{e:cslgw}
\end{equation}
where the Gutzwiller projector
\begin{equation}
\textrm{P}_{\scriptscriptstyle\textrm{GW}} \equiv \prod_{i=1}^N
\big(1-c^\dagger_{i\up}c_{i\up}\, c^\dagger_{i\dw}c_{i\dw} \big)
\label{e:gwp}
\end{equation}
eliminates doubly occupied sites and 
$\ket{\psi^{N}_{\scriptscriptstyle\textrm{SD}}}$ 
is the Slater determinant wave function for the lowest Landau level 
filled once with $M=\frac{N}{2}$ $\up$-spin and once with 
$M$ $\dw$-spin electrons.

The spin singlet property of the chiral spin liquid mentioned above is 
now evident.  Clearly the Slater determinant
$\ket{\psi^{N}_{\scriptscriptstyle\textrm{SD}}}$ is a singlet.  Since the
Gutzwiller projector commutes with the spin operator on each site,
\begin{equation}
\comm{\textrm{P}_{\scriptscriptstyle\textrm{GW}}}{\boldsymbol{S}_i} = 0,
\label{e:comm}
\end{equation}
$\ket{\psi^{\scriptscriptstyle\textrm{CSL}}_+}$ must be a singlet as well.

The Gutzwiller form (\ref{e:cslgw}) also allows a rather elegant formulation
of the spinon excitations.  For example, a state with $L$ $\dw$-spin spinons
at sites $\eta_1,\ldots,\eta_L$ is given by
\begin{equation}
\ket{\psi^{\scriptscriptstyle\textrm{CSL}}_{+\eta_1\dw\ldots \eta_L\dw }}
\,=\,\textrm{P}_{\scriptscriptstyle\textrm{GW}} 
\,c_{\eta_1\up}\ldots c_{\eta_L\up}\,
\ket{\psi^{N+L}_{\scriptscriptstyle\textrm{SD}}},
\label{e:cslgwsp}
\end{equation}
where $N+L=2M$ must be an even integer.
The equivalence of (\ref{e:cslgwsp}) to (\ref{e:psichisp}) is readily
seen with (\ref{e:zw}).  This form nicely illustrates the (fractional)
spin \half of the spinon.  The electron annihilation operators
create inhomogeneities in spin and charge before projection.  The 
projector enforces one particle per site and hence restores the 
homogeneity in the charge distribution, but commutes with spin.
We are left with a neutral object of spin \half.

Note that the spinon coordinates $\eta$ in either (\ref{e:psichisp})
or (\ref{e:cslgwsp}) do not have to coincide with lattice points.
This allows us to define a winding path and thus a statistical
parameter via a Berry's phase encountered upon adiabatic interchange.

Holon excitations, which carry a positive unit charge and no spin,
are constructed from spinon excitations by annihilation 
of an electron at the spinon site, which must now coincide with a 
lattice point.  For example, a chiral spin liquid with $L$ holons
at sites $\eta_1,\ldots,\eta_L$ is given by
\begin{equation}
\ket{\psi^{\scriptscriptstyle\textrm{CSL}}_{+\eta_1\circ\ldots \eta_L\circ }}
\,=\,c_{\eta_1\dw}\ldots c_{\eta_L\dw}\,
\textrm{P}_{\scriptscriptstyle\textrm{GW}} 
\,c_{\eta_1\up}\ldots c_{\eta_L\up}\,
\ket{\psi^{N+L}_{\scriptscriptstyle\textrm{SD}}},
\label{e:cslgwho}
\end{equation}
where $N+L=2M$ must again be an even integer.

The chiral spin liquid, when formulated as a Gutzwiller projection of
filled Landau levels, is readily generalized to different lattice
types.

\section{THE SPIN 1 CHIRALITY LIQUID}
\label{s:cl}

The spin 1 chirality liquid is obtained by combining 
$\psi^{\scriptscriptstyle\textrm{CSL}}_+$ and its complex conjugate 
$\psi^{\scriptscriptstyle\textrm{CSL}}_-$ via a projection to
spin 1 on each site, which can be accomplished elegantly via 
Schwinger bosons.  Let 
\begin{equation}
\begin{array}{rcl}\displaystyle
\ket{\psi^{\scriptscriptstyle\textrm{CSL}}_+}
&=&\displaystyle\sum_{\scriptstyle\{z_1,\ldots ,z_M; w_1,\ldots,w_{M} \}}\,
\phi (z_1,\ldots,z_M)\,\phi (w_1,\ldots,w_{M})\,\cdot\rule{15pt}{0pt} 
\\ &&\displaystyle\rule{110pt}{0pt}\displaystyle
\cdot\,a^\dagger_{z_1}\ldots a^\dagger_{z_M}\,
       b^\dagger_{w_1}\ldots b^\dagger_{w_M}\vac = \\ 
&\equiv&\displaystyle \Psi^{\scriptscriptstyle\textrm{CSL}}_+
\big[a^\dagger ,b^\dagger\big]\,\vac .\rule{0pt}{20pt}
\end{array}
\label{e:cslschw}
\end{equation}
be a chiral spin liquid (\ref{e:csl}) written in terms of
Schwinger bosons.  The \mbox{spin 1} chirality liquid 
is then given by\\[-1pt]
\begin{equation}
\ket{\psi^{\scriptscriptstyle\textrm{S1CL}}}
\;=\;
\Psi^{\scriptscriptstyle\textrm{CSL}}_+\big[a^\dagger ,b^\dagger\big]\;
\Psi^{\scriptscriptstyle\textrm{CSL}}_-\big[a^\dagger ,b^\dagger\big]\,
\vac .
\label{e:cl}
\end{equation}\\[-1pt]
This is the central proposal of this article.

The spinon or holon excitations for the spin 1 chirality liquid 
are just the spinon or holon excitations
of the individual chiral spin liquids.  
As the spinons for the chiral spin liquid are massive, we expect 
an energy gap for the spin 1 liquid as well.
The spinons in the latter carry a chirality quantum
number $\chi$, which is $+1$ if they are constructed as
excitations of $\psi^{\scriptscriptstyle\textrm{CSL}}_+$ and $-1$
if they are constructed as
excitations of $\psi^{\scriptscriptstyle\textrm{CSL}}_-$. 

Let us first consider the case where we only have spinons of chirality
$+$.  They span a Hilbert space of dimension $2^N$, just as in the case 
of the chiral spin liquid or Haldane-Shastry chain.  The spinons
obey a fractional exclusion principle according to half fermi statistics.
The adiabatic transport argument of Arovas, Schrieffer, and 
Wilczek\cite{arov} remains unaffected by the projection as well.
We can evaluate the phase acquired by 
$\Psi^{\scriptscriptstyle\textrm{CSL}}_+\big[a^\dagger ,b^\dagger\big] $
and hence $\ket{\psi^{\scriptscriptstyle\textrm{S1CL}}}$ as we
exchange two $+$ chirality spinons 
by winding them counterclockwise around each other, and obtain $\pi/2$.
If we had only spinons of chirality $-$, we would obtain $-\pi/2$.
In this sense, spinons of either chirality obey half-fermi statistics, 
with statistical parameter 
\begin{equation}
\theta=\chi\frac{\pi}{2}.  
\label{e:clstat}
\end{equation}

Non-trivialities arise, 
however, as we consider states with spinons 
of both chiralities.  If spinons of different chiralities were
independent of each other, the total Hilbert space dimension of the
ground state plus all possible spinon excitations would be $2^N
\!\cdot\! 2^N = 4^N$.  The Hilbert space dimension for a spin 1 system
with $N$ sites, however, is only $3^N$.  The spinon basis is 
consequently overcomplete, and spinons of different chiralities cannot
always be orthogonal.  They have a significant overlap when they are
spatially close to each other.

What does this mean?  One possibility is that a parameterization of
the spin 1 Hilbert space in terms of a ground state plus spinons 
excitations is not sensible.  This would be the case if there is no 
energy scale at which spinon degrees of freedom offer an adequate 
description of the system.  I do not believe in this possibility
but cannot rule it out at this stage.

The other possibility is that there is a way to transform the $4^N$ 
non-orthogonal orbitals into $3^N$ orthogonal orbitals, which then 
provide the desired Hilbert space basis.  Note that
a similar situation had occurred in the Haldane-Shastry model, when
we first constructed spinon excitations in real space and 
ended up with a vastly overcomplete basis.  The problem was
resolved by a transformation to momentum space.  Of course, 
the possibility of such a transformation 
confronts us with a variety of unresolved questions.  The
most pressing one, in my opinion, concerns the statistics.
Since we do not expect the transformation to dissolve the concept
of spinons, it will be linear and hence preserve the fractional 
exclusion principle.  It is not clear, however, what happens to 
chirality quantum numbers, winding phases, and relative angular momenta.

Fortunately, the issue of overcompleteness only matters when we 
have spinons of different chiralities nearby.  For the low energy
dynamics of a spin 1 chirality liquid another question is
more immediate:  are the spinons free, as they are in spin \half
chains or in chiral spin liquids, or are they confined, as they 
are in spin 1 chains?  Even though I will not be able to give a 
definite answer, I will present a modest amount of 
evidence for a conjecture in section \ref{s:con}%
\ \  First, however, I will interpret 
a few numerical studies on the 
the spin 1 chirality liquid ground state in the following section.

\section{FINITE SIZE STUDIES} 
\label{s:num}

In this section, we report on a few numerical studies of the 
spin 1 chirality liquid (S1CL) for square, triangular and 
kagom\'e lattices, 
as well as the chiral spin liquids (CSLs) used to construct them.  
The calculations are performed on finite size systems with periodic 
boundary conditions (PBC).  To construct the spin liquid states 
for this geometry, we have to generalize the wave function
(\ref{e:ll}) for a filled Landau level to PBCs, which is in essence
done by replacing the factors $(z_j-z_k)$ by odd Jacobi theta
functions
$\vartheta_{\scriptscriptstyle\frac{1}{2},\frac{1}{2}}
(z_j-z_k,\tau )$.\cite{hr}

\vspace{3pt}
The most severe limitation for exact diagonalization studies is the
size of the Hilbert space.  For a spin 1 system, the dimension of the
subspace with $S^z_{\textrm{tot}}=0$ is given by
\begin{equation}
\sum_{n=0}^{N/2} \frac{N!}{n!\, n!\, (N-2n)!} 
\end{equation}
which amounts to 5,196,627 for N=16.  This limits the system size to
an order of 16 sites, which is far too small to address most questions
of interest.  For example, it is well established that all the spin 1
Heisenberg antiferromagnets we study possess long range order.  This
feature, however, is hardly manifest in the systems we consider here.
The comparisons of the spin liquid trial wave functions with exact
ground states of Heisenberg models reported in Table \ref{t:num}
merely indicate to which extent the energetically relevant
nearest-neighbor correlations of the models are captured by the
liquids.  If these correlations are sufficiently close, we may
conjecture that the spin liquid is stabilized if we disorder the
system by doping.

\begin{table}
\begin{center}
\begin{tabular}{|c|c|c|c|c|c|} \hline
\multicolumn{2}{|c|}{Lattice type} 
  &\multicolumn{2}{|c|}{square} &triangular &kagom\'e\rule{0pt}{13pt} \\[3pt]
\cline{3-4}\multicolumn{2}{|c|}{} 
&$J'\!=\!0$ &$J'\!=\!\frac{1}{2}J$ & &\rule{0pt}{12pt} \\[3pt]\hline
\multicolumn{2}{|c|}{Number of sites $N$}     
  &\multicolumn{2}{|c|}{16}   &16     &12\rule{0pt}{13pt}\\[3pt]
\multicolumn{2}{|c|}{Principal displacement $\boldsymbol{L_1}$ } 
&\multicolumn{2}{|c|}{(4,0)}  &(4,0)  &(4,0)           \\[3pt]
\multicolumn{2}{|c|}{\phantom{Principal displacement} $\boldsymbol{L_2}$}
&\multicolumn{2}{|c|}{(0,4)}         &(2,$2\sqrt{3}$) &(2,$2\sqrt{3}$)\\[3pt]
\multicolumn{2}{|c|}{LL bound.\ phases $\varphi_1$,$\varphi_2$}
&\multicolumn{2}{|c|}{$\pi$,$\pi$} &$\frac{2\pi}{3}$,$\frac{4\pi}{3}$ 
&$\frac{\pi}{4}$,$\frac{3\pi}{4}$\\[3pt]\hline
                &$E_{\scriptscriptstyle\textrm{exact}}$ 
                  &-11.2285 &-8.4579 &-8.5555 &-5.4448\rule{0pt}{13pt}\\[3pt]
$S=\frac{1}{2}$ &$E_{\scriptscriptstyle\textrm{CSL}}$   
                  & -8.2572 &-7.6435 &-7.8251 &-4.9756\\[3pt]
                &$\braket{\psi^{\scriptscriptstyle\textrm{CSL}}|
                  \psi^{\scriptscriptstyle\textrm{exact}}}$  
                  &.6727 &.8626 &.0000 &.0000 \\[3pt]\hline
                &$E_{\scriptscriptstyle\textrm{exact}}$ 
                  &-38.3047 &-25.0858 &-25.3699 &-17.6208
                  \rule{0pt}{13pt} \\[3pt] \cline{2-6}
                  \rule{0pt}{13pt}
                &$E_{\scriptscriptstyle\textrm{S1CL}}$   
                  &-30.7987 &-24.4782 &-24.1504 &-15.7025\\[3pt]
                & $\Delta E/E_{\scriptscriptstyle\textrm{exact}}$
                  &19.6\,\% &2.42\,\% &4.80\,\% &10.89\,\% \\[3pt]
\raisebox{2pt}{$S=1$}           
                &$\braket{\psi^{\scriptscriptstyle\textrm{S1CL}}|
                  \psi^{\scriptscriptstyle\textrm{exact}}}$ 
                  &.6019 &.9484 &.7948 &.5527 \\[3pt] \cline{2-6}
                  \rule{0pt}{13pt}
                &$E_{\scriptscriptstyle\textrm{S1F1/2}}$ 
                  &-38.1286 &-24.8957 &-25.0184 &-15.9249\\[3pt]
                & $\Delta E/E_{\scriptscriptstyle\textrm{exact}}$
                  &0.46\,\% &0.76\,\% &1.39\,\% &9.62\,\% \\[3pt]
                &$\braket{\psi^{\scriptscriptstyle\textrm{S1F1/2}}|
                  \psi^{\scriptscriptstyle\textrm{exact}}}$ 
                  &.9927 &.9784 &.8962 &.02954 \\[3pt]\hline
\end{tabular}
\end{center}
\caption{Energy expectation values and overlaps 
for the chiral spin liquid (CSL), the spin 1 chirality liquid (S1CL),
and a spin 1 liquid constructed by combining two identical copies of
the exact spin \half ground states according to the Schwinger boson 
projection principle discussed in this article (S1F1/2), for finite lattices
of various types with PBC.  The trial states are compared to 
exact ground states of nearest-neighbor Heisenberg models with $J=1$
for all three lattice types, plus a model augmented by a 
next-nearest-neighbor coupling $J'$ on the square lattice.
The displacement vectors 
$\boldsymbol{L_1}$, $\boldsymbol{L_2}$ span the principal region for 
the PBCs in units of lattice constants.  The boundary phases 
$\varphi_1$ and $\varphi_2$ are a property of the Landau level wave functions
adapted to PBCs\cite{hr} and projected onto the lattice, 
which are used to construct the chiral spin
liquids:  $\varphi_1$ and $\varphi_2$ are the phases acquired by
$\phi_{\scriptscriptstyle\boldsymbol{L_1},\boldsymbol{L_2}} 
(z_1,\ldots,z_M)$ as one of the coordinates 
(which must coincide with a lattice point) is translated by 
$\boldsymbol{L_1}$ or $\boldsymbol{L_2}$, respectively, plus the phase
acquired by a unit charge under this translation through the coupling to
the vector potential generating the magnetic field.\cite{subtle}  
These phases have been chosen to minimize the energies
of the corresponding spin liquids, which slightly depend on it.
}
\label{t:num}
\end{table}

Let us now take a look at Table \ref{t:num}.  First of all, 
the results confirm
that the P and T violating CSLs are
inadequate for all three lattice types.  The energies always differ by
an amount of order 10\,\% or larger, which is enormous in light of the
fact that the energy of the classical ground state is only about 30\,\%
higher than the energy of the quantum model.  Let us now turn to the
S1CL for the square lattice.  For a Heisenberg antiferromagnet with
only a nearest-neighbor coupling $J$, the S1CL is is way off as well,
as shown in the first column of Table \ref{t:num}.  If we use,
however, the S1CL as a trial wave function for a Heisenberg model with
a next-nearest-neighbor coupling $J'\!=\!J/2$, we find an energy
difference of 2.42\,\% and an overlap of 95\,\%, which is reasonable.
There is a very realistic chance that the S1CL is realized in this
antiferromagnet if it is disordered by additional frustration or
disorder.  It may further be a candidate for a disordered state on the
triangular lattice but appears inadequate for the kagom\'e lattice.

In the last three lines on the bottom of Table \ref{t:num}, I have
evaluated energies and overlaps of yet another trial wave function for
$s=1$ antiferromagnets, which is simply obtained by combining two
identical copies of the exact ground states for the corresponding
$s=\frac{1}{2}$ Heisenberg model
\begin{equation}
\ket{\psi^{\scriptscriptstyle\textrm{exact}}}
\,=\,\Psi^{\scriptscriptstyle\textrm{1/2}}
\big[a^\dagger ,b^\dagger\big]\,\vac 
\label{e:exhalf}
\end{equation}
by Schwinger boson projection:
\begin{equation}
\ket{\psi^{\scriptscriptstyle\textrm{S1F1/2}}}
\;=\;
\Psi^{\scriptscriptstyle\textrm{1/2}}\big[a^\dagger ,b^\dagger\big]\;
\Psi^{\scriptscriptstyle\textrm{1/2}}\big[a^\dagger ,b^\dagger\big]\,
\vac .
\label{e:sonefhalf}
\end{equation}
I call this trial state ``\,spin 1 from 1/2\,'' (S1F1/2).  Comparisons
of this state with the exact ground states of the spin 1
antiferromagnets reveal information about how different the correlations
in the spin 1 and spin \half models are.  For the
nearest-neighbor Heisenberg model on the square lattice, the energy of
these states differs only by 0.46\,\%, with an overlap of 99\,\%.
I attribute this to the fact that both states share the same type of long
range order.  For the triangular lattice, this overlap is already
reduced to 90\,\%.  I interpret this as an indication that the order
is much more pronounced in the triangular spin 1 system.  For the kagom\'e
lattice, the overlap is virtually zero, which confirms that
spin \half and spin 1 models possess very different correlations.  The
spin \half system is an intrinsically disordered spin
liquid\cite{kag}, while the spin 1 model on the kagom\'e lattice
possesses order.  Most likely, there is a disordered state for the spin 
1 kagom\'e system which can be stabilized by doping, but it is
apparently not obtained by Schwinger boson projection of two
identical spin \half liquids.  I conjecture that nature avails
herself of the possibility that the individual spin \half states used
in this construction may violate a discrete symmetry, which is
not P or T, but symmetry under rotations by $\pi/3$ 
modulo $2\pi/3$.  This is, however, merely a speculation.

\section{SPINON CONFINEMENT} 
\label{s:con}

In section \ref{s:aklt}, we illustrated with a little cartoon how a
linear confinement force between spinons arises for the AKLT chain,
even though spinons are free for the individual Majumdar-Gosh chains.
This example might suggest that if spinons are confined, this property
can always be understood through such a simple picture.  Let me first
demonstrate that this is not the case, and then turn to the subtle
question of whether spinons are confined in the spin 1 chirality
liquid.

Consider a spin 1 Gutzwiller chain (S1GW), that is, a spin 1 chain
obtained by Schwinger boson projection of two identical Haldane-Shastry 
chains:
\begin{equation}
\ket{\psi^{\scriptscriptstyle\textrm{S1GW}}}
\;=\;
\Psi^{\scriptscriptstyle\textrm{HS}}_{\scriptscriptstyle 0}
\big[a^\dagger ,b^\dagger\big]\;
\Psi^{\scriptscriptstyle\textrm{HS}}_{\scriptscriptstyle 0} 
\big[a^\dagger ,b^\dagger\big]\,
\vac .
\label{e:sonehs}
\end{equation}
As the individual Haldane-Shastry chains are completely disordered,
there is no indication of confining forces on the level of the cartoon
mentioned above.  Nonetheless, the S1GW chain does not support free
spinons.  There are several ways to see this.  The first is that the
S1GW chain is in the same universality class as the AKLT or spin 1
Heisenberg chain, which is known to possess an energy gap\cite{sonehs}.  
Since the spinons for the Haldane-Shastry are (in contrast to the
Majumdar-Gosh chain or the chiral spin liquid) massless, the energy gap
must stem from spinon-spinon interactions, or confining forces.  This
is in accordance with Haldane's analysis\cite{hgap,affl} of the
one-dimensional nonlinear \mbox{$\sigma$ model,} which shows that a
topological term and hence deconfined spinons exist only for half-integer
spin chains.

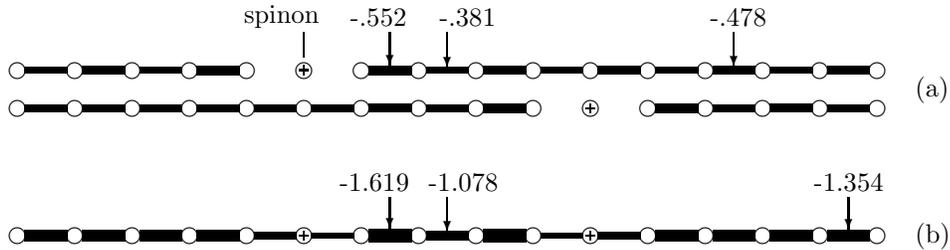
\begin{figure}
\begin{center}
\setlength{\unitlength}{0.240900pt}

\begin{picture}(1440,400)(0,-10)
\font\ef=cmr8 
\font\tf=cmr10 
\tf
\put(  45,   0){{\makebox(0,0){\rule{15.90pt}{ 4.06pt}}}}
\put(   0,   0){\circle{25}}
\put(1350,   0){\circle{25}}
\put( 135,   0){{\makebox(0,0){\rule{15.90pt}{ 4.15pt}}}}
\put(  90,   0){\circle{25}}
\put( 225,   0){{\makebox(0,0){\rule{15.90pt}{ 3.94pt}}}}
\put( 180,   0){\circle{25}}
\put( 315,   0){{\makebox(0,0){\rule{15.90pt}{ 4.39pt}}}}
\put( 270,   0){\circle{25}}
\put( 405,   0){{\makebox(0,0){\rule{15.90pt}{ 2.66pt}}}}
\put( 360,   0){\circle{25}}
\put( 495,   0){{\makebox(0,0){\rule{15.90pt}{ 1.57pt}}}}
\put( 450,   0){\circle{25}}
\put( 585,   0){{\makebox(0,0){\rule{15.90pt}{ 4.86pt}}}}
\put( 540,   0){\circle{25}}
\put( 675,   0){{\makebox(0,0){\rule{15.90pt}{ 3.23pt}}}}
\put( 630,   0){\circle{25}}
\put( 765,   0){{\makebox(0,0){\rule{15.90pt}{ 4.86pt}}}}
\put( 720,   0){\circle{25}}
\put( 855,   0){{\makebox(0,0){\rule{15.90pt}{ 1.57pt}}}}
\put( 810,   0){\circle{25}}
\put( 945,   0){{\makebox(0,0){\rule{15.90pt}{ 2.66pt}}}}
\put( 900,   0){\circle{25}}
\put(1035,   0){{\makebox(0,0){\rule{15.90pt}{ 4.39pt}}}}
\put( 990,   0){\circle{25}}
\put(1125,   0){{\makebox(0,0){\rule{15.90pt}{ 3.94pt}}}}
\put(1080,   0){\circle{25}}
\put(1215,   0){{\makebox(0,0){\rule{15.90pt}{ 4.15pt}}}}
\put(1170,   0){\circle{25}}
\put(1305,   0){{\makebox(0,0){\rule{15.90pt}{ 4.06pt}}}}
\put(1260,   0){\circle{25}}
\put( 450,   0){{\makebox(0,0){\rule{ 3.40pt}{  .40pt}}}}
\put( 450,   0){{\makebox(0,0){\rule{  .40pt}{ 3.40pt}}}}
\put( 900,   0){{\makebox(0,0){\rule{ 3.40pt}{  .40pt}}}}
\put( 900,   0){{\makebox(0,0){\rule{  .40pt}{ 3.40pt}}}}

\put( 560,  70){\makebox(0,0)[b]{-1.619}}
\put( 585,  60){\vector(0,-1){ 49}}
\put( 700,  70){\makebox(0,0)[b]{-1.078}}
\put( 675,  60){\vector(0,-1){ 53}}
\put(1305,  70){\makebox(0,0)[b]{-1.354}}
\put(1305,  60){\vector(0,-1){ 51}}
\put(1410,   0){\makebox(0,0)[l]{(b)}}
\put(  45, 260){{\makebox(0,0){\rule{15.90pt}{ 2.44pt}}}}
\put(   0, 260){\circle{25}}
\put(1350, 260){\circle{25}}
\put( 135, 260){{\makebox(0,0){\rule{15.90pt}{ 2.96pt}}}}
\put(  90, 260){\circle{25}}
\put( 225, 260){{\makebox(0,0){\rule{15.90pt}{ 2.29pt}}}}
\put( 180, 260){\circle{25}}
\put( 315, 260){{\makebox(0,0){\rule{15.90pt}{ 3.31pt}}}}
\put( 270, 260){\circle{25}}
\put( 405, 260){{\makebox(0,0){\rule{15.90pt}{  .00pt}}}}
\put( 360, 260){\circle{25}}
\put( 495, 260){{\makebox(0,0){\rule{15.90pt}{  .00pt}}}}
\put( 450, 260){\circle{25}}
\put( 585, 260){{\makebox(0,0){\rule{15.90pt}{ 3.31pt}}}}
\put( 540, 260){\circle{25}}
\put( 675, 260){{\makebox(0,0){\rule{15.90pt}{ 2.29pt}}}}
\put( 630, 260){\circle{25}}
\put( 765, 260){{\makebox(0,0){\rule{15.90pt}{ 2.96pt}}}}
\put( 720, 260){\circle{25}}
\put( 855, 260){{\makebox(0,0){\rule{15.90pt}{ 2.44pt}}}}
\put( 810, 260){\circle{25}}
\put( 945, 260){{\makebox(0,0){\rule{15.90pt}{ 2.89pt}}}}
\put( 900, 260){\circle{25}}
\put(1035, 260){{\makebox(0,0){\rule{15.90pt}{ 2.48pt}}}}
\put( 990, 260){\circle{25}}
\put(1125, 260){{\makebox(0,0){\rule{15.90pt}{ 2.87pt}}}}
\put(1080, 260){\circle{25}}
\put(1215, 260){{\makebox(0,0){\rule{15.90pt}{ 2.48pt}}}}
\put(1170, 260){\circle{25}}
\put(1305, 260){{\makebox(0,0){\rule{15.90pt}{ 2.89pt}}}}
\put(1260, 260){\circle{25}}
\put( 450, 260){{\makebox(0,0){\rule{ 3.40pt}{  .40pt}}}}
\put( 450, 260){{\makebox(0,0){\rule{  .40pt}{ 3.40pt}}}}

\put( 560, 330){\makebox(0,0)[b]{ -.552}}
\put( 585, 320){\vector(0,-1){ 53}}
\put( 700, 330){\makebox(0,0)[b]{ -.381}}
\put( 675, 320){\vector(0,-1){ 55}}
\put(1125, 330){\makebox(0,0)[b]{ -.478}}
\put(1125, 320){\vector(0,-1){ 54}}
\put( 415, 327){\makebox(0,0)[b]{spinon}}
\put( 450, 320){\line(0,-1){ 40}}
\put(1410, 230){\makebox(0,0)[l]{(a)}}
\put(  45, 200){{\makebox(0,0){\rule{15.90pt}{ 2.89pt}}}}
\put(   0, 200){\circle{25}}
\put(1350, 200){\circle{25}}
\put( 135, 200){{\makebox(0,0){\rule{15.90pt}{ 2.48pt}}}}
\put(  90, 200){\circle{25}}
\put( 225, 200){{\makebox(0,0){\rule{15.90pt}{ 2.87pt}}}}
\put( 180, 200){\circle{25}}
\put( 315, 200){{\makebox(0,0){\rule{15.90pt}{ 2.48pt}}}}
\put( 270, 200){\circle{25}}
\put( 405, 200){{\makebox(0,0){\rule{15.90pt}{ 2.89pt}}}}
\put( 360, 200){\circle{25}}
\put( 495, 200){{\makebox(0,0){\rule{15.90pt}{ 2.44pt}}}}
\put( 450, 200){\circle{25}}
\put( 585, 200){{\makebox(0,0){\rule{15.90pt}{ 2.96pt}}}}
\put( 540, 200){\circle{25}}
\put( 675, 200){{\makebox(0,0){\rule{15.90pt}{ 2.29pt}}}}
\put( 630, 200){\circle{25}}
\put( 765, 200){{\makebox(0,0){\rule{15.90pt}{ 3.31pt}}}}
\put( 720, 200){\circle{25}}
\put( 855, 200){{\makebox(0,0){\rule{15.90pt}{  .00pt}}}}
\put( 810, 200){\circle{25}}
\put( 945, 200){{\makebox(0,0){\rule{15.90pt}{  .00pt}}}}
\put( 900, 200){\circle{25}}
\put(1035, 200){{\makebox(0,0){\rule{15.90pt}{ 3.31pt}}}}
\put( 990, 200){\circle{25}}
\put(1125, 200){{\makebox(0,0){\rule{15.90pt}{ 2.29pt}}}}
\put(1080, 200){\circle{25}}
\put(1215, 200){{\makebox(0,0){\rule{15.90pt}{ 2.96pt}}}}
\put(1170, 200){\circle{25}}
\put(1305, 200){{\makebox(0,0){\rule{15.90pt}{ 2.44pt}}}}
\put(1260, 200){\circle{25}}
\put( 900, 200){{\makebox(0,0){\rule{ 3.40pt}{  .40pt}}}}
\put( 900, 200){{\makebox(0,0){\rule{  .40pt}{ 3.40pt}}}}
\end{picture}
\end{center}
\caption{Spin-spin correlations in two spin \half Haldane-Shastry chains 
  with 15 sites and one spinon each (a) and a spin 1 Gutzwiller chain (b)
  constructed via Schwinger boson projection from two Haldane-Shastry
  chains with spinons at different sites.  The thickness of the lines
  reflects the magnitudes of the antiferromagnetic correlations
  $\braket{\boldsymbol{S}_i \boldsymbol{S}_{i+1}}$ on nearest neighbor
  sites.}
\label{f:chn}
\end{figure}
The second way to see that spinons in the S1GW chain are confined is
by direct computation of the confining potential, as shown in
\mbox{Fig.~\ref{f:chn}.}  The numerically accessible system size is
again to small for a full analysis, as we must consider periodic
boundary conditions\cite{open} and are hence limited to a few lattice
spacings when we pull two spinons apart.  Nonetheless,
Fig.~\ref{f:chn} clearly shows a dimerization of the spin correlations
between the spinons.  The dimerized chain segment compares unfavorably
in energy to the uniformly correlated chain far away.  I should
mention here that when I say spinons are confined, I do not
necessarily mean that there is a string force between them, as it is
the case for spinons in AKLT chains or two-leg Heisenberg
ladders\cite{ladder}.  I merely mean that it costs less energy to
create another spinon pair than to completely separate two spinons.
In the S1GW chain, the confining force approaches a constant for large
spinon separations, as a single localized spinon in a Haldane-Shastry
chain dimerizes the spin correlations over the entire chain.  The
strength of the dimerization and hence the force between widely
separated spinons, however, decreases with the system size, and
vanishes in the thermodynamic limit.\cite{mimo}

Let us now turn to the question of whether there is spinon confinement
in the spin 1 chirality liquid.  First of all, from an analysis of 
the nonlinear $\sigma$ model in two dimensions, there is no reason
to expect deconfined spinons for either half-integer or integer
spins.\cite{nohopf} The fact that the chiral spin liquid supports
deconfined spinons may be interpreted as yet another indication that
it does not describe the universality class of the generic spin \half
liquid in two dimensions, which is believed to be of central
importance to the problem of CuO superconductivity.

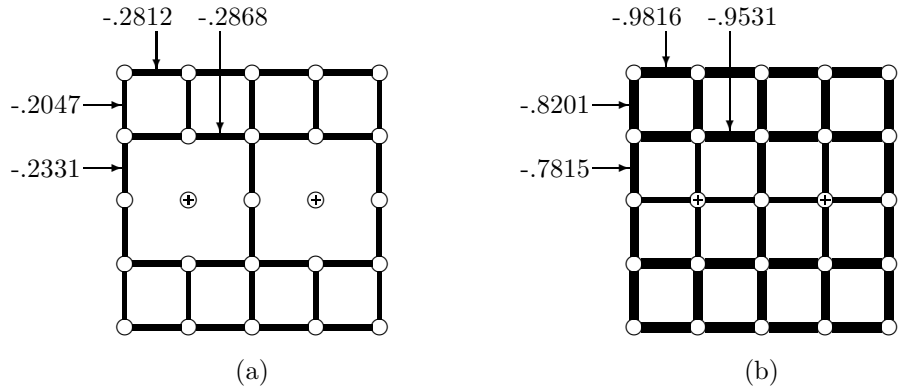
\begin{figure}
\begin{center}
\setlength{\unitlength}{0.240900pt}
\begin{picture}(1500,600)(-230,-270)
\font\ef=cmr8 
\font\tf=cmr10 
\tf
\put( 800,  50){{\makebox(0,0){\rule{ 3.13pt}{18.31pt}}}}
\put(1200,  50){{\makebox(0,0){\rule{ 3.13pt}{18.31pt}}}}
\put( 850,   0){{\makebox(0,0){\rule{18.31pt}{ 1.89pt}}}}
\put( 800,   0){\circle{25}}
\put(1200,   0){\circle{25}}
\put( 800, 150){{\makebox(0,0){\rule{ 3.28pt}{18.31pt}}}}
\put(1200, 150){{\makebox(0,0){\rule{ 3.28pt}{18.31pt}}}}
\put( 850, 100){{\makebox(0,0){\rule{18.31pt}{ 3.81pt}}}}
\put( 800, 100){\circle{25}}
\put(1200, 100){\circle{25}}
\put( 800,-150){{\makebox(0,0){\rule{ 3.28pt}{18.31pt}}}}
\put(1200,-150){{\makebox(0,0){\rule{ 3.28pt}{18.31pt}}}}
\put( 850,-200){{\makebox(0,0){\rule{18.31pt}{ 3.93pt}}}}
\put( 850, 200){{\makebox(0,0){\rule{18.31pt}{ 3.93pt}}}}
\put( 800,-200){\circle{25}}
\put(1200,-200){\circle{25}}
\put( 800, 200){\circle{25}}
\put(1200, 200){\circle{25}}
\put( 800, -50){{\makebox(0,0){\rule{ 3.13pt}{18.31pt}}}}
\put(1200, -50){{\makebox(0,0){\rule{ 3.13pt}{18.31pt}}}}
\put( 850,-100){{\makebox(0,0){\rule{18.31pt}{ 3.81pt}}}}
\put( 800,-100){\circle{25}}
\put(1200,-100){\circle{25}}
\put( 900,  50){{\makebox(0,0){\rule{ 2.03pt}{18.31pt}}}}
\put( 950,   0){{\makebox(0,0){\rule{18.31pt}{ 1.89pt}}}}
\put( 900,   0){\circle{25}}
\put( 900, 150){{\makebox(0,0){\rule{ 3.24pt}{18.31pt}}}}
\put( 950, 100){{\makebox(0,0){\rule{18.31pt}{ 3.81pt}}}}
\put( 900, 100){\circle{25}}
\put( 900,-150){{\makebox(0,0){\rule{ 3.24pt}{18.31pt}}}}
\put( 950,-200){{\makebox(0,0){\rule{18.31pt}{ 3.93pt}}}}
\put( 950, 200){{\makebox(0,0){\rule{18.31pt}{ 3.93pt}}}}
\put( 900,-200){\circle{25}}
\put( 900, 200){\circle{25}}
\put( 900, -50){{\makebox(0,0){\rule{ 2.03pt}{18.31pt}}}}
\put( 950,-100){{\makebox(0,0){\rule{18.31pt}{ 3.81pt}}}}
\put( 900,-100){\circle{25}}
\put(1000,  50){{\makebox(0,0){\rule{ 3.13pt}{18.31pt}}}}
\put(1050,   0){{\makebox(0,0){\rule{18.31pt}{ 1.89pt}}}}
\put(1000,   0){\circle{25}}
\put(1000, 150){{\makebox(0,0){\rule{ 3.28pt}{18.31pt}}}}
\put(1050, 100){{\makebox(0,0){\rule{18.31pt}{ 3.81pt}}}}
\put(1000, 100){\circle{25}}
\put(1000,-150){{\makebox(0,0){\rule{ 3.28pt}{18.31pt}}}}
\put(1050,-200){{\makebox(0,0){\rule{18.31pt}{ 3.93pt}}}}
\put(1050, 200){{\makebox(0,0){\rule{18.31pt}{ 3.93pt}}}}
\put(1000,-200){\circle{25}}
\put(1000, 200){\circle{25}}
\put(1000, -50){{\makebox(0,0){\rule{ 3.13pt}{18.31pt}}}}
\put(1050,-100){{\makebox(0,0){\rule{18.31pt}{ 3.81pt}}}}
\put(1000,-100){\circle{25}}
\put(1100,  50){{\makebox(0,0){\rule{ 2.03pt}{18.31pt}}}}
\put(1150,   0){{\makebox(0,0){\rule{18.31pt}{ 1.89pt}}}}
\put(1100,   0){\circle{25}}
\put(1100, 150){{\makebox(0,0){\rule{ 3.24pt}{18.31pt}}}}
\put(1150, 100){{\makebox(0,0){\rule{18.31pt}{ 3.81pt}}}}
\put(1100, 100){\circle{25}}
\put(1100,-150){{\makebox(0,0){\rule{ 3.24pt}{18.31pt}}}}
\put(1150,-200){{\makebox(0,0){\rule{18.31pt}{ 3.93pt}}}}
\put(1150, 200){{\makebox(0,0){\rule{18.31pt}{ 3.93pt}}}}
\put(1100,-200){\circle{25}}
\put(1100, 200){\circle{25}}
\put(1100, -50){{\makebox(0,0){\rule{ 2.03pt}{18.31pt}}}}
\put(1150,-100){{\makebox(0,0){\rule{18.31pt}{ 3.81pt}}}}
\put(1100,-100){\circle{25}}
\put( 900,   0){{\makebox(0,0){\rule{ 3.40pt}{  .40pt}}}}
\put( 900,   0){{\makebox(0,0){\rule{  .40pt}{ 3.40pt}}}}
\put(1100,   0){{\makebox(0,0){\rule{ 3.40pt}{  .40pt}}}}
\put(1100,   0){{\makebox(0,0){\rule{  .40pt}{ 3.40pt}}}}
\put( 730,  50){\makebox(0,0)[r]{-.7815}}
\put( 730, 150){\makebox(0,0)[r]{-.8201}}
\put( 735,  50){\vector(1,0){ 58}}
\put( 735, 150){\vector(1,0){ 58}}
\put( 820, 275){\makebox(0,0)[b]{-.9816}}
\put( 970, 275){\makebox(0,0)[b]{-.9531}}
\put( 850, 265){\vector(0,-1){ 56}}
\put( 950, 265){\vector(0,-1){157}}
\put(1000,-250){\makebox(0,0)[t]{(b)}}
\put(   0,  50){{\makebox(0,0){\rule{ 1.86pt}{18.31pt}}}}
\put( 400,  50){{\makebox(0,0){\rule{ 1.86pt}{18.31pt}}}}
\put(  50,   0){{\makebox(0,0){\rule{18.31pt}{  .00pt}}}}
\put(   0,   0){\circle{25}}
\put( 400,   0){\circle{25}}
\put(   0, 150){{\makebox(0,0){\rule{ 1.64pt}{18.31pt}}}}
\put( 400, 150){{\makebox(0,0){\rule{ 1.64pt}{18.31pt}}}}
\put(  50, 100){{\makebox(0,0){\rule{18.31pt}{ 2.29pt}}}}
\put(   0, 100){\circle{25}}
\put( 400, 100){\circle{25}}
\put(   0,-150){{\makebox(0,0){\rule{ 1.64pt}{18.31pt}}}}
\put( 400,-150){{\makebox(0,0){\rule{ 1.64pt}{18.31pt}}}}
\put(  50,-200){{\makebox(0,0){\rule{18.31pt}{ 2.25pt}}}}
\put(  50, 200){{\makebox(0,0){\rule{18.31pt}{ 2.25pt}}}}
\put(   0,-200){\circle{25}}
\put( 400,-200){\circle{25}}
\put(   0, 200){\circle{25}}
\put( 400, 200){\circle{25}}
\put(   0, -50){{\makebox(0,0){\rule{ 1.86pt}{18.31pt}}}}
\put( 400, -50){{\makebox(0,0){\rule{ 1.86pt}{18.31pt}}}}
\put(  50,-100){{\makebox(0,0){\rule{18.31pt}{ 2.29pt}}}}
\put(   0,-100){\circle{25}}
\put( 400,-100){\circle{25}}
\put( 100,  50){{\makebox(0,0){\rule{  .00pt}{18.31pt}}}}
\put( 150,   0){{\makebox(0,0){\rule{18.31pt}{  .00pt}}}}
\put( 100,   0){\circle{25}}
\put( 100, 150){{\makebox(0,0){\rule{ 1.69pt}{18.31pt}}}}
\put( 150, 100){{\makebox(0,0){\rule{18.31pt}{ 2.29pt}}}}
\put( 100, 100){\circle{25}}
\put( 100,-150){{\makebox(0,0){\rule{ 1.69pt}{18.31pt}}}}
\put( 150,-200){{\makebox(0,0){\rule{18.31pt}{ 2.25pt}}}}
\put( 150, 200){{\makebox(0,0){\rule{18.31pt}{ 2.25pt}}}}
\put( 100,-200){\circle{25}}
\put( 100, 200){\circle{25}}
\put( 100, -50){{\makebox(0,0){\rule{  .00pt}{18.31pt}}}}
\put( 150,-100){{\makebox(0,0){\rule{18.31pt}{ 2.29pt}}}}
\put( 100,-100){\circle{25}}
\put( 200,  50){{\makebox(0,0){\rule{ 1.86pt}{18.31pt}}}}
\put( 250,   0){{\makebox(0,0){\rule{18.31pt}{  .00pt}}}}
\put( 200,   0){\circle{25}}
\put( 200, 150){{\makebox(0,0){\rule{ 1.64pt}{18.31pt}}}}
\put( 250, 100){{\makebox(0,0){\rule{18.31pt}{ 2.29pt}}}}
\put( 200, 100){\circle{25}}
\put( 200,-150){{\makebox(0,0){\rule{ 1.64pt}{18.31pt}}}}
\put( 250,-200){{\makebox(0,0){\rule{18.31pt}{ 2.25pt}}}}
\put( 250, 200){{\makebox(0,0){\rule{18.31pt}{ 2.25pt}}}}
\put( 200,-200){\circle{25}}
\put( 200, 200){\circle{25}}
\put( 200, -50){{\makebox(0,0){\rule{ 1.86pt}{18.31pt}}}}
\put( 250,-100){{\makebox(0,0){\rule{18.31pt}{ 2.29pt}}}}
\put( 200,-100){\circle{25}}
\put( 300,  50){{\makebox(0,0){\rule{  .00pt}{18.31pt}}}}
\put( 350,   0){{\makebox(0,0){\rule{18.31pt}{  .00pt}}}}
\put( 300,   0){\circle{25}}
\put( 300, 150){{\makebox(0,0){\rule{ 1.69pt}{18.31pt}}}}
\put( 350, 100){{\makebox(0,0){\rule{18.31pt}{ 2.29pt}}}}
\put( 300, 100){\circle{25}}
\put( 300,-150){{\makebox(0,0){\rule{ 1.69pt}{18.31pt}}}}
\put( 350,-200){{\makebox(0,0){\rule{18.31pt}{ 2.25pt}}}}
\put( 350, 200){{\makebox(0,0){\rule{18.31pt}{ 2.25pt}}}}
\put( 300,-200){\circle{25}}
\put( 300, 200){\circle{25}}
\put( 300, -50){{\makebox(0,0){\rule{  .00pt}{18.31pt}}}}
\put( 350,-100){{\makebox(0,0){\rule{18.31pt}{ 2.29pt}}}}
\put( 300,-100){\circle{25}}
\put( 100,   0){{\makebox(0,0){\rule{ 3.40pt}{  .40pt}}}}
\put( 100,   0){{\makebox(0,0){\rule{  .40pt}{ 3.40pt}}}}
\put( 300,   0){{\makebox(0,0){\rule{ 3.40pt}{  .40pt}}}}
\put( 300,   0){{\makebox(0,0){\rule{  .40pt}{ 3.40pt}}}}
\put( -70,  50){\makebox(0,0)[r]{-.2331}}
\put( -70, 150){\makebox(0,0)[r]{-.2047}}
\put( -65,  50){\vector(1,0){ 61}}
\put( -65, 150){\vector(1,0){ 61}}
\put(  20, 275){\makebox(0,0)[b]{-.2812}}
\put( 170, 275){\makebox(0,0)[b]{-.2868}}
\put(  50, 265){\vector(0,-1){ 60}}
\put( 150, 265){\vector(0,-1){160}}
\put( 200,-250){\makebox(0,0)[t]{(a)}}
\end{picture}
\end{center}
\caption{Spin-spin correlations in a chiral spin liquid (a) and a
  spin 1 chirality liquid (b) constructed with it for a 16 site square
  lattice with periodic boundary conditions in the presence of two
  spinons at the positions indicated.}
\label{f:sq}
\end{figure}
Evidence for confining forces in the spin 1 chirality liquid is
presented in Fig.~\ref{f:sq}.  In a CSL with two spinons (a), the
energetically relevant nearest-neighbor spin correlations are slightly
enhanced near the spinons.  This enhancement is due to the fact that
the sites around the spinons effectively have fewer neighbors, as
there is no need to correlate the links to the spinon sites.
Combining this $+$ chirality CSL with a $-$ chirality CSL ground state
(which has spatially uniform correlations) yields a S1CL with two $+$
chirality spinons (b).  The spin correlations for this state are
slightly depressed in the immediate vicinity of the spinons, in sharp
contrast to what one would expect from the correlations in the CSLs.
I interpret this depression as an indication of confining forces
between the spinons.  I should emphasize, however, that the system
size is too small to allow for a conclusion.

If one believes that there is a confining force between spinons in 
the S1CL, one is led to wonder where it is coming form.  In the
spin 1 chain, we have been able to trace it to a dimerization
in the spin correlations in the chain segment between the spinons.
The spin correlations provided information about both the existence
and the origin of the force.  In the S1CL, however, Fig.~\ref{f:sq}
only suggests that a force exists, but does not reveal its origin.
It seems to come from no-where.
The origin of the force, I believe, only reveals itself 
if one considers chirality correlations, that is, expectation values
of the chirality operator\cite{wwz} 
\begin{equation}
\chi\,\equiv\,\boldsymbol{S}_i\cdot
(\boldsymbol{S}_{j}\times \boldsymbol{S}_{k}),
\label{e:chiop}
\end{equation}
where $i$,$j$, and $k$ are three lattice sites on the same plaquet.
This operator has been used as an order parameter for the CSL, as it
effectively provides a local measure of the P and T violation in a
spin liquid.  The expectation value is trivially zero in the S1CL
ground state.   As the individual spinons in the S1CL carry chirality
and hence violate P and T, we expect a cloud of chirality, that is,
non-zero expectation values of (\ref{e:chiop}) on triangles,
around an isolated spinon.  This chirality upsets the spin
correlations, and costs energy.  The chiral disturbance can be repaired
by a second spinon nearby, which according to my intuition could be of
either chirality.  The confining force can hence be traced back to the
appearance of chirality in the liquid as one pulls spinons apart.  I
conjecture that similar forces play a central role in the pairing of
charge carriers in CuO superconductors.\cite{else}

\section{CONCLUSION}
\label{s:conc}

There are many interesting and unresolved questions regarding the generic 
spin 1 liquid state in two dimensions.  The spin 1 chirality liquid 
introduced here offers a paradigm for this state and hence provides
a framework to formulate, and possibly resolve, some of these.

\section*{ACKNOWLEDGMENTS}

This research was supported through NSF grant DMR-95-21888 while
at Stanford University, and through Habilitationsstipendium Gr
1715/1-1 of the Deutsche Forschungsgemeinschaft while at the
University of Karlsruhe.  I wish to thank T.\ Kopp for his critical
reading of the manuscript.



\begin{thebibliography}{99}
\bibitem{sutwo} I implicitly assume the spin group is SU(2).
\bibitem{foot1} 
    Strictly speaking, this proof, which is in principle
    due Majumdar and Gosh\cite{mg} (see also ref.~\onlinecite{auer}),
    only shows that the states (\ref{e:psimg}) are ground states, but
    not that they are the only ground states of (\ref{e:hmg}).  
    It is,
    however, rather simple to complete the proof.  Marshall\cite{mars}
    has shown in his classic treatment of antiferromagnetism that any
    singlet configuration of a spin-$\frac{1}{2}$ system can be
    written as a sum over dimer configurations, where the dimers are
    taken over all possible pairs of spins.  (The space of all
    possible pairings is much larger then the space of singlets, so that
    the mapping is not unique.)  It is easy to see that none of these
    dimer configurations except for those given in (\ref{e:psimg}) are
    simultaneously annihilated by $H_i$ for all $i$.  The argument is
    easily generalized to triplet and higher total spin states.
\bibitem{foot3} Note that these spinons are only approximate
  eigenstates of $H^{\scriptscriptstyle\textrm{MG}}$.
\bibitem{unfair} This view does not reflect the history adequately.
  D.H.\ Lee thought of using the quantized Hall wave function for
  bosons with $m=2$ to formulate a spin liquid, and Kalmeyer and
  Laughlin\cite{chiral} showed that it is a singlet when multiplied by
  a gauge factor $G(z_j)$.  The Haldane-Shastry model was discovered
  only in the following year.
\bibitem{subtle} For the kagom\'e lattice, these phases are not (and
  cannot be) equal for all lattice sites.  The chiral spin liquid on
  the kagom\'e lattice specified in Table \ref{t:num} is not invariant
  under rotations.
\bibitem{sonehs} Numerically, the S1GW chain is is an even better
  approximation to the spin 1 nearest-neighbor Heisenberg chain than
  the AKLT chain.  For a 10 site chain with PBCs, the energies are
  2.60\,\% and 5.39\,\% off, respectively.  The overlaps with the
  exact ground states are 0.9450 and 0.8638.
\bibitem{open} The problem with open boundary conditions is that the
  generic spin 1 chain has spinons at the ends.  This can easily be
  seen from an AKLT cartoon.
\end{thebibliography}
\end{document}